\documentclass[lettersize,journal]{IEEEtran}
\usepackage{cite}
\usepackage{amsmath,amssymb,amsfonts}
\usepackage{algorithmic}
\usepackage{graphicx}
\usepackage{textcomp}
\def\BibTeX{{\rm B\kern-.05em{\sc i\kern-.025em b}\kern-.08em
    T\kern-.1667em\lower.7ex\hbox{E}\kern-.125emX}}
\usepackage{booktabs}
\usepackage[table,xcdraw]{xcolor}
\usepackage{epsfig}
\usepackage{epstopdf}
\usepackage[nolist]{acronym}
\usepackage{mathtools}
\usepackage{siunitx}
\DeclareSIUnit{\belmilliwatt}{Bm}
\DeclareSIUnit{\dBm}{\deci\belmilliwatt}

\usepackage{algorithmic}
\usepackage{algorithm}
\usepackage{enumitem}
\usepackage{hyperref}
\usepackage{subcaption} 
\usepackage{multirow}
\usepackage{printlen}

\usepackage{amsthm}

\begin{acronym}
	\acro{MIMO}{multiple-input multiple-output}
	\acro{SNR}{signal-to-noise ratio}
	\acro{CP}{cyclic prefix}
	\acro{OFDM}{orthogonal frequency division multiplexing}
	\acro{CSI}{channel state information}
	\acro{PDP}{power delay profile}
	\acro{DFT}{discrete Fourier transform}
	\acro{ICI}{inter-carrier interference}
	\acro{OTFS}{orthogonal time frequency space}
	\acro{AWGN}{additive white Gaussian noise}
	\acro{QAM}{quadrature amplitude modulation}
	\acro{ISI}{inter-symbol interference}
	\acro{ISAC}{integrated sensing and communication}
	\acro{AoA}{angle of arrival}
	\acro{ToA}{time of arrival}
	\acro{CPI}{coherent processing interval}
	\acro{CFO}{carrier frequency offset}
	\acro{ULA}{uniform linear antenna array}
	\acro{CRB}{Cram\'er-Rao bound}
	\acro{LoS}{line of sight}
	\acro{AoD}{angle of departure}
	\acro{ML}{machine learning}
	\acro{BS}{base station}
	\acro{LS}{least squares}
	\acro{MUSIC}{multiple signal classification}
	\acro{SVD}{singular value decomposition}
	\acro{MSE}{mean squared error}
	\acro{RCS}{radar cross-section}
	\acro{IoT}{Internet of things}
	\acro{UAV}{unmanned aerial vehicle}
	\acro{FIM}{Fisher information matrix}
	\acro{CIR}{channel impulse response}
	\acro{TO}{timing offset}
	\acro{FO}{frequency offset}
	\acro{FFT}{fast Fourier transform}
	\acro{IFFT}{inverse fast Fourier transform}
	\acro{MLP}{multi-layer perceptron}
	\acro{NN}{neural network}
	\acro{DL}{deep learning}
	\acro{MLE}{maximum likelihood estimator}
	\acro{TDD}{time-division duplexing}
	\acro{mmWave}{milli-meter wave}
	\acro{C2VNN}{convolutional complex-valued neural network}
    \acro{SAGE}{space-alternating generalized expectation-maximization}
    \acro{ESPRIT}{estimation of signal parameters via rotational invariant techniques}
    \acro{CNN}{convolutional neural network}
    \acro{FNN}{feedforward neural network}
    \acro{GPU}{graphics processing unit}
    \acro{FLOPS}{floating point operations per second}
    \acro{FPGA}{field programmable gate array}
    \acro{ppm}{parts per million}
    \acro{MAE}{mean absolute error}
    \acro{MMSE}{minimum mean squared error}
    \acro{RMSE}{root mean squared error}
    \acro{IDFT}{inverse discrete Fourier transform}
    \acro{DML}{deterministic maximum likelihood}
    \acro{HPC}{high-performance computing}
    \acro{HPBW}{half-power beamwidth}
    \acro{PAC}{probably approximately correct}
    \acro{THz}{terahertz}
    \acro{RIS}{reconfigurable intelligent surface}
    \acro{TDoA}{time difference of arrival}
	\acro{PDF}{probability density function}    
	\acro{RF}{radio frequency}
	\acro{CORDIC}{coordinate rotation digital computer}
\end{acronym}

\DeclareMathOperator{\diag}{diag}
\DeclareMathOperator{\CRB}{CRB}
\DeclareMathOperator{\vectorize}{vec}
\DeclareMathOperator{\TSVD}{TSVD}

\DeclareMathOperator{\subjectto}{subject \ to}

\DeclareMathOperator*{\argmin}{arg\,min}

\newcommand{\CLASSINPUTtoptextmargin}{1.8cm}

\usepackage{soul,color}

\usepackage{geometry}
\usepackage{titlesec}

\titlespacing{\subsection}{3pt}{*0.05}{*0.1}
\titlespacing{\subsubsection}{3pt}{*0.05}{*0.1}
\usepackage{fancyhdr}
\fancypagestyle{firststyle}{
\fancyhf{}
\fancyhead[L]{S. Naoumi, A. Bazzi, R. Bomfin, M. Chafii, "High-Resolution Sensing in Communication-Centric ISAC: Deep Learning and Parametric Methods", accepted in \textit{IEEE Journal on Selected Areas in Communications}, 2025.}

}

\begin{document}

\title{High-Resolution Sensing in Communication-Centric ISAC: Deep Learning and Parametric Methods}

\author{Salmane Naoumi, Ahmad Bazzi, Roberto Bomfin, Marwa Chafii 
\thanks{
This work is supported in part by the NYUAD Center for Artificial Intelligence and Robotics, funded by Tamkeen under the Research Institute Award CG010. This research was carried out on the High Performance Computing resources at New York University Abu Dhabi.

Salmane Naoumi is with NYU Tandon School of Engineering, Brooklyn, 11201, NY, USA (email: sn3397@nyu.edu).

Roberto Bomfin is with the Engineering Division, New York University (NYU) Abu Dhabi, UAE.

Ahmad Bazzi and Marwa Chafii are with the Engineering Division, New York University (NYU) Abu Dhabi, UAE and NYU WIRELESS, NYU Tandon School of Engineering, Brooklyn, NY. 

}
}

\maketitle
\thispagestyle{firststyle}
  
\IEEEpubid{}

\begin{abstract}
This paper introduces two novel algorithms designed to address the challenge of super-resolution sensing parameter estimation in bistatic configurations within communication-centric integrated sensing and communication (ISAC) systems. Our approach leverages the estimated channel state information derived from reference symbols originally intended for communication to achieve super-resolution sensing parameter estimation. The first algorithm, IFFT-C2VNN, employs complex-valued convolutional neural networks to estimate the parameters of different targets, achieving significant reductions in computational complexity compared to traditional methods. The second algorithm, PARAMING, utilizes a parametric method that capitalizes on the knowledge of the system model, including the transmit and receive array geometries, to extract the sensing parameters accurately. Through a comprehensive performance analysis, we demonstrate the effectiveness and robustness of both algorithms across a range of signal-to-noise ratios, underscoring their applicability in realistic ISAC scenarios.
\end{abstract}


\section{Introduction}
\label{sec:introduction}
\Ac{ISAC} has been identified as a foundational technology expected to shape the future of 6G wireless systems \cite{LiuJSAC}. This emerging paradigm enables the joint integration of radar sensing and communication functionalities within a unified framework, thereby enhancing spectrum efficiency and reducing both hardware and computational costs \cite{10018908}. Such advancements open transformative possibilities across a wide range of applications, including automotive technology, \ac{IoT} \cite{9206051}, and robotics \cite{chafii2023twelve}.

The \ac{ISAC} framework can be classified into three primary areas of research: communication-centric, radar-centric, and joint design \cite{9540344, 10038611}. In radar-centric approaches, communication data is embedded within radar waveforms \cite{radar-centric}, whereas joint design methods focus on the co-optimization of sensing and communication systems, balancing these functionalities to meet application-specific requirements \cite{8386661}. Recently, communication-centric \ac{ISAC} has emerged as an effective approach to augment existing communication infrastructure with sensing capabilities. This approach leverages transmitted communication waveforms to extract sensing parameters, such as range and velocity, of targets within the environment. Studies have demonstrated the efficacy of communication waveforms for sensing, particularly in configurations such as mono-static \ac{ISAC} systems that employ \ac{OTFS} modulation \cite{10279367} and \ac{TDD} massive \ac{MIMO} systems \cite{10250200}.
Among communication-centric configurations, the bistatic setting has gained recognition as one of the most practical solutions. By employing separate transmit and receive antennas, the bistatic configuration integrates seamlessly with existing communication infrastructures while expanding sensing coverage and minimizing interference \cite{perf_ca_s}. Extensive research on bistatic radar systems has shown their effectiveness in estimating environmental sensing parameters by repurposing the \ac{CSI} estimation process \cite{close_paper_R2, 10496165}. Furthermore, recent studies have demonstrated the suitability of \ac{OFDM}-based waveforms in bistatic \ac{ISAC} setups, further underscoring the feasibility of this configuration for practical implementations \cite{close_paper_R1, Bomfin_UW}.

In bistatic \ac{ISAC} systems, a central challenge is super-resolution estimation of \ac{AoD}, \ac{AoA}, time-of-flight, and Doppler. Super-resolution is needed to approach the sub-meter localization targets envisioned for 6G \cite{parameter_estim}, yet it often relies on fitting high-order parametric models that exceed the resolution limits of conventional processing. The resulting computational burden complicates real-time deployment under practical hardware and latency constraints, and performance is tightly coupled to the granularity of the search space, which further increases cost \cite{framework_pmn}.
Subspace methods such as \ac{MUSIC} and \ac{ESPRIT} mitigate complexity by exploiting signal and noise subspaces for parameter recovery \cite{music_isac,esprit_isac}. Their accuracy, however, typically requires many coherent samples and degrades in low \ac{SNR} or when scatterers are closely spaced. Compressive approaches cast estimation as a sparse recovery problem and leverage structure across space, time, and frequency to achieve super-resolution with fewer measurements \cite{joint_compressive_sensing_2}. As the problem dimension grows, these techniques incur rapidly increasing computational cost, and their measurement demands are difficult to satisfy in short-coherence channels and with limited array sizes \cite{close_paper_R3}.

Beyond classical grid scans, a substantial body of state-of-the-art research has advanced multi-dimensional super-resolution for joint parameter retrieval. Multilinear subspace and tensor formulations such as $R$-D ESPRIT and Tensor-ESPRIT exploit shift invariances across space, frequency, and time to estimate angles, delays, and Doppler with polynomial-time algebraic solvers while mitigating off-grid bias \cite{8003323}. A recent unified tensor framework for massive \ac{MIMO} \ac{ISAC} further enables joint channel and target estimation with favorable identifiability under limited snapshots \cite{tensor_unified_isac_2024}. In \ac{OFDM}-centric \ac{ISAC}, frequency-selective coupling across tones has been explicitly modeled to derive estimators and performance limits for angle-delay-Doppler recovery in the presence of \ac{ICI} \cite{KeskinWymeersch_JSTSP2021_ICI}.

Complementary gridless sparse approaches based on atomic norms and sparse Bayesian learning provide off-grid super-resolution with reduced bias and strong low-\ac{SNR} behavior, including blind and semi-blind formulations for \ac{MIMO}-\ac{OFDM} that jointly recover \ac{AoA}, \ac{AoD}, and delay \cite{10509713,10328457,8537983}. Related progress in mmWave FD-\ac{MIMO} has yielded parametric estimators such as multi-dimensional unitary ESPRIT that are tailored to wideband arrays and hardware impairments, and that enable accurate joint angle-delay recovery with modest pilot overhead \cite{8846224}. In parallel, optimization-driven receivers based on lifted atomic norms and on joint target-data detection have been proposed for \ac{ISAC}, and they demonstrate simultaneous recovery of locations, velocities, delays, and communication symbols under realistic signaling and bistatic operation \cite{lanm_isac,jtd_bistatic_isac}. Very recent efforts further refine tensor and subspace formulations for joint sensing-communication inference in \ac{ISAC}, which strengthens identifiability insights and informs algorithmic design \cite{PLAIN_2025}.

Orthogonal lines of work broaden the operating regimes of \ac{ISAC}. At \ac{THz} frequencies, massive-\ac{MIMO} \ac{ISAC} leverages channel training and tensor decompositions under hybrid architectures to scale joint channel and target estimation \cite{10643882}. Carrier-aggregation designs fuse low and high bands with \ac{OFDM} pilots and compressed sensing in order to enhance range-velocity resolution across multi-band resources \cite{10285442}. In addition, \ac{RIS}-aided mmWave \ac{ISAC} optimizes reflections and beamforming to minimize the \ac{CRB}, which highlights the accuracy gains achievable through propagation reconfiguration \cite{10422722}. Finally, \ac{OTFS}-based designs leverage delay-Doppler sparsity to improve range-velocity estimation and robustness to channel dynamics, and therefore provide an attractive alternative to \ac{OFDM}-domain processing \cite{otfs_isac_spie,10941856}. In parallel, advances in \ac{ML}, and in particular \ac{DL}, offer computationally efficient super-resolution through data-driven priors and improved robustness to model mismatches \cite{prev_paper_22,prev_paper_18}. Recent surveys document the growing role of \ac{NN}-based estimators in radar signal processing, with gains in resolution and generalization across array sizes and hardware imperfections \cite{surveySa}. Within \ac{ISAC}, \ac{DL} has proved effective in vehicular settings and has been adapted to dual parameter estimation in uplink \ac{OFDM} systems \cite{isac_dl_applications_2,isac_dl_applications_3}. To the best of our knowledge, however, learning-based joint estimation tailored to bistatic communication-centric \ac{ISAC} remains largely unexplored, which motivates the developments presented in this work.

In this paper, we address this gap by proposing a novel \ac{DL}-based model for joint sensing parameter estimation in bistatic communication-centric \ac{ISAC} setups. Specifically, we focus on the estimation of \ac{AoA}, \ac{AoD}, and \ac{ToA} parameters. In addition, we propose a parametric method that leverages the system characteristics as well as the structure of the estimated channel matrix to estimate the sensing parameters. Both methods aim to achieve high estimation accuracy while significantly reducing computational complexity, thus enabling real-time applications within next-generation wireless systems.

In summary, our work makes the following key contributions
\begin{itemize}
    \item We introduce \textit{IFFT-\ac{C2VNN}}, a specialized \ac{DL} architecture tailored for high-resolution estimation of sensing parameters. The proposed method leverages coarse estimates obtained from the \ac{IFFT} of the estimated channel matrix to focus computational resources on regions of interest around target peaks in the transformed domain. This preprocessing enables IFFT-\ac{C2VNN} to efficiently capture the fine-grained details necessary for precise parameter estimation. The model architecture integrates complex-valued convolutional layers and enhances the estimation precision with minimal computational overhead. Training is performed using simulation data under varying \ac{SNR} conditions, with \ac{MSE} employed as the training loss function.
    
    \item We propose \textit{PARAMING}, a PARAmetric method for joint angles and tiMING estimation that exploits the full space-time structure of the system model. More precisely, PARAMING restructures the estimated \ac{CSI} into compact sub-array matrices, making full use of array geometry, as well as \ac{OFDM} structure. A truncated \ac{SVD} is then applied to isolate the principal components of the transformed matrix, enabling accurate \ac{ToA} estimation for each target/clutter component without requiring a grid search. Subsequently, a two-stage \ac{LS} fitting process followed by 2D regression is applied to jointly compute the \ac{AoA} and \ac{AoD} estimates for each \ac{ToA} value, thereby providing fine-grained spatial and temporal super-resolution. 
    As a result, PARAMING provides 3D sensing (\ac{AoA}, \ac{AoD} and \ac{ToA}) information for each target and clutter component with low complexity and high resolution, by leveraging model-based transformations.

    \item We present a comprehensive computational complexity analysis of PARAMING and IFFT-\ac{C2VNN}, quantifying the required multiplications and additions for each method. Additionally, we compare the proposed methods to the conventional \ac{MLE} approach, highlighting their significant computational advantages, particularly for real-time processing and high-resolution sensing tasks.
    
    \item We conduct a comprehensive evaluation of the proposed PARAMING and IFFT-\ac{C2VNN} methods for estimating key sensing parameters, comparing their performance against state-of-the-art methods. The results demonstrate the superior estimation accuracy and robustness of the proposed methods across varying \ac{SNR} levels. Furthermore, we show that both methods achieve significantly lower latency compared to grid-based approaches, making them highly efficient for real-time \ac{ISAC} applications. Additionally, we extend the proposed methods to include Doppler frequency estimation, demonstrating their adaptability and strong performance in scenarios with moving targets.
\end{itemize}
\textbf{Notation}: Upper-case and lower-case boldface letters denote matrices and vectors, resp. $(\cdot)^T$, $(\cdot)^*$ and $(\cdot)^H$ represent the transpose, the conjugate and the transpose-conjugate operators.
We denote by $*$ the convolution operator. 
For any complex number $z \in \mathbb{C}$,
the real part of $z$ is denoted as $\Re(z)$, whereas the imaginary part is denoted as $\Im(z)$. The $\ell_2$ norm of a vector $\pmb{x}$ is denoted as $\Vert \pmb{x} \Vert$. The matrices  $\pmb{F}$ and $\pmb{I}$ are the Fourier and the identity matrices with appropriate dimensions, resp. 
For matrix indexing, the $(i,j)^{th}$ entry of matrix $\pmb{A}$ is denoted by $[\pmb{A}]_{[i,j]}$ and its $j^{th}$ column is denoted as $\pmb{A}_{[:,j]}$. The operator $\otimes$ is the \textit{Kronecker} product.
The big-$\mathcal{O}$ notation is $\mathcal{O}()$.
For a set $\mathcal{A} = \lbrace a_1 \ldots a_N \rbrace$ containing integers, the notation $\mathcal{A} + k$ adds integer $k$ to the elements of $\mathcal{A}$, i.e. $\mathcal{A} + k = \lbrace a_1 + k \ldots a_N + k \rbrace$.

\section{System Model}
\label{sec:system_model}
\begin{figure}[!t]
    \centering
    \includegraphics[scale=0.41]{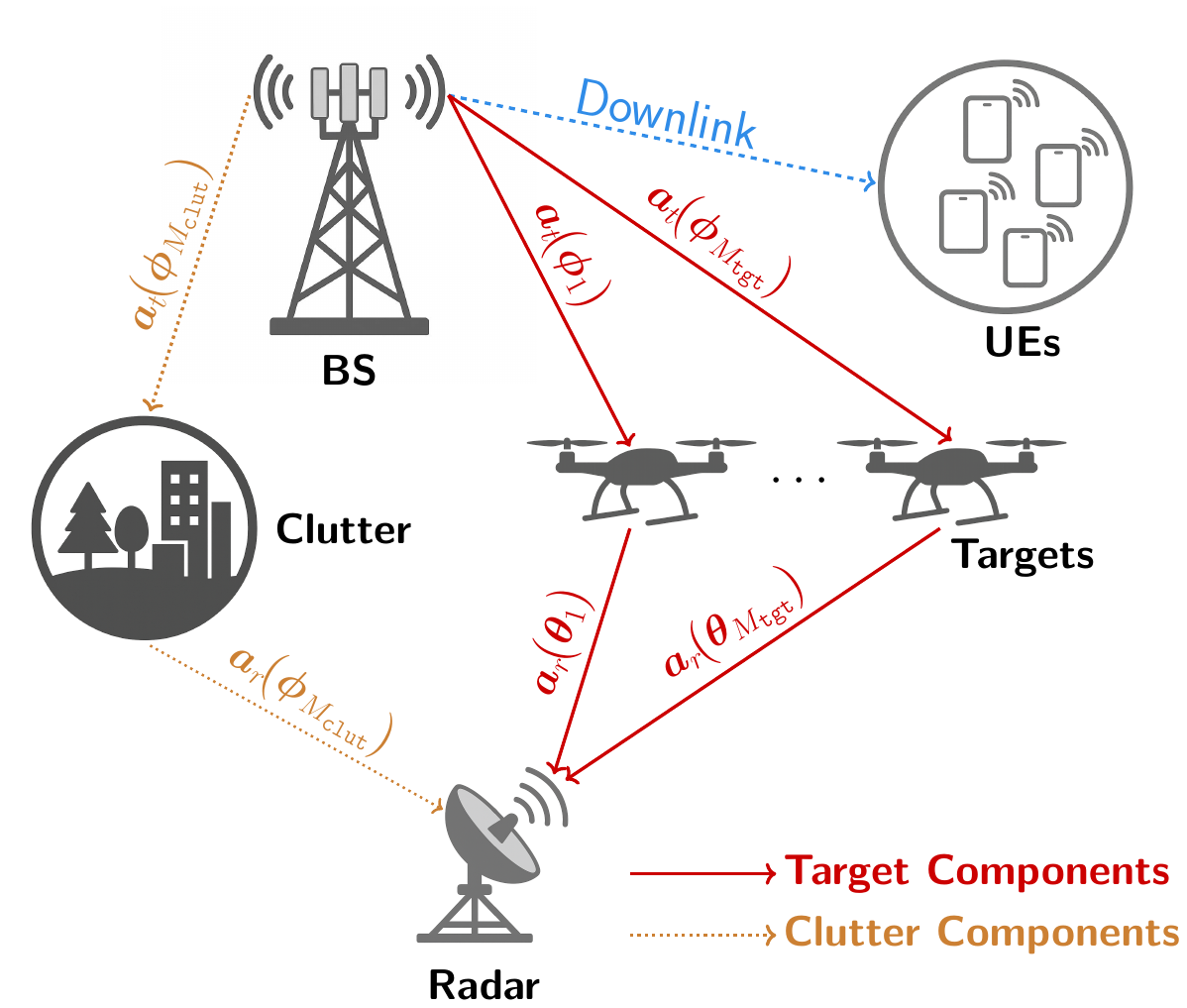}
    \caption{Illustration of the bistatic communication-centric \ac{ISAC} system model, where downlink transmissions from a \ac{BS} to communication users serve as simultaneous illumination sources for radar sensing at a passive radar unit.}
    \label{fig:system_model}
\end{figure}
This section introduces the communication-centric \ac{ISAC} framework considered in this study, in which the downlink transmission from a \ac{BS} to communication users also serves as an illumination source for passive radar sensing. This system configuration, illustrated in Fig.~\ref{fig:system_model}, supports joint sensing and communication functionalities by leveraging transmitted communication signals for environmental sensing.
\subsection{Transmitted Signal Model}
The \ac{BS} is equipped with a \ac{ULA} comprising \(N_{\tt{t}}\) antennas and transmits frames of \(K_{\tt{P}}\) \ac{OFDM} symbols over \(N_{\tt{P}}\) active subcarriers. Each \ac{OFDM} symbol has a duration \(T = \frac{1}{\Delta_f}\), where \(\Delta_f\) is the subcarrier spacing. 
To mitigate inter-symbol and inter-carrier interference, a \ac{CP} of duration \(T_{\tt{CP}}\) is appended to each symbol; unless otherwise stated, \(T_{\tt{CP}}\) is chosen to exceed the maximum excess delay \(\tau_{\max}\) of the environment. The total symbol duration is \(T_o = T + T_{\tt{CP}}\).
The transmitted signal for the \(k^{\text{th}}\) \ac{OFDM} symbol within a frame is
\begin{equation}
    \label{eq:tx-OFDM-DL}
    \pmb{x}_{k}(t) = \sum_{n=1}^{N_{\tt{P}}} \pmb{s}_{n,k} \, c_n(-t) \, \Pi(t - k T_o), \quad k = 1, \ldots, K_{\tt{P}},
\end{equation}
where \(\Pi(t)\) denotes a rectangular window function of length \(T_o\), \(c_n(\tau) = e^{-j 2 \pi n \Delta_f \tau}\) is the delay response associated with the \(n^{\text{th}}\) subcarrier, and \(\pmb{s}_{n,k} \in \mathbb{C}^{N_{\tt{t}} \times 1}\) represents the modulated symbol on the \(n^{\text{th}}\) subcarrier for the \(k^{\text{th}}\) \ac{OFDM} symbol.
\subsection{Sensing Channel Model}
\label{subsec:sensing_channel}
The sensing environment contains multiple specular contributions from both targets and environmental clutter. Let
\(
M \!=\! M_{\tt{tgt}} \!+\! M_{\tt{clut}}
\)
denote the total number of resolvable propagation paths. Each path \(m\in\{1,\ldots,M\}\) is parameterized by the \ac{AoA} \(\theta_m\), \ac{AoD} \(\phi_m\), \ac{ToA} \(\tau_m\), and Doppler shift \(f_{D,m}\).
The continuous-time \ac{CIR} between the \(n_t^{\text{th}}\) BS transmit element and the \(n_r^{\text{th}}\) radar receive element is
\begin{equation}
\label{eq:CIR}
\begin{split}
    h_{n_r,n_t}(t,\tau) 
    = &
    \sum\limits_{m=1}^{M}
    \alpha_m(t) e^{j 2 \pi \big( f_{D,m} + f_{\tt{off}} \big) t}
    a_{n_r}(\theta_m)
    a_{n_t}(\phi_m)  
    \\ &
    \delta \big(\tau - \tau_m - \tau_{\tt{off}}\big),
\end{split}
\end{equation}
\noindent where \(a_{n_r}(\theta_m)\) and \(a_{n_t}(\phi_m)\) are the receive and transmit steering coefficients, \(\delta(\cdot)\) is the Dirac delta function, and \((\tau_{\tt off},f_{\tt off})\) capture both \acp{TO} and \acp{FO} due to imperfect synchronization between the \ac{BS} and radar unit.

Within a \ac{CPI}, the complex gain $\alpha_m(t)$ is assumed slowly varying, i.e., $\alpha_m(t)\!\approx\!\alpha_m$. We model $\alpha_m\in\mathbb{C}$ as an aggregate per-path factor that captures propagation loss, \ac{RF} gains and the bistatic \ac{RCS}
\begin{equation}
\label{eq:alpha_factor_local_paper}
\alpha_m
\;=\;
\sqrt{G_{\tt t}(\phi_m)\,G_{\tt r}(\theta_m)}\;
\frac{\lambda\,\sqrt{\sigma^{\tt bi}_m}}{(4\pi)^{3/2}\,R_{{\tt t},m}\,R_{{\tt r},m}}
\;e^{j\delta_m},
\end{equation}
where $\sigma^{\tt bi}_m$ denotes the bistatic \ac{RCS}, $R_{{\tt t},m}$ and $R_{{\tt r},m}$ are the \ac{BS} to target and target to radar ranges, $\lambda$ is the wavelength, and $\delta_m$ accounts for residual phase terms. The functions $G_{\tt t}(\cdot)$ and $G_{\tt r}(\cdot)$ denote the transmit and receive antenna gains, respectively.

While the spatially separated bistatic configuration is suitable for \ac{ISAC} applications, it introduces asynchronous clock issues that may impair the sensing accuracy through measurement ambiguities and time-varying phase shifts, limiting coherent processing across time slots \cite{async_rheath_paper}.
In what follows, we consider the near-perfect synchronization regime (i.e., \(\tau_{\tt off}\!\approx\!0\), \(f_{\tt off}\!\approx\!0\)); otherwise these terms can be estimated/compensated using standard or advanced techniques \cite{chen2023kalman_4, jump, 9349171_31}.

The frequency-domain channel at the \(n^{\text{th}}\) subcarrier and \(k^{\text{th}}\) \ac{OFDM} symbol is given by
\begin{equation}
\label{eq:CIR-freq-doppler}
    \pmb{H}_{n,k}
    \;=\;
    \pmb{A}_r(\pmb{\Theta})\;
    \pmb{G}_k(\pmb{f_D})\;
    \pmb{D}_n(\pmb{\tau})\;
    \pmb{A}_t^{\!T}(\pmb{\Phi})
    \;\in\; \mathbb{C}^{N_{\tt r}\times N_{\tt t}},
\end{equation}
\noindent with steering matrices
\(
\pmb{A}_t(\pmb{\Phi})=\big[\,\pmb{a}_t(\phi_1)\; \cdots\; \pmb{a}_t(\phi_M)\,\big]
\),
\(
\pmb{A}_r(\pmb{\Theta})=\big[\,\pmb{a}_r(\theta_1)\; \cdots\; \pmb{a}_r(\theta_M)\,\big]
\),
per-path Doppler/complex gains
\(
\pmb{G}_k(\pmb{f_{D}}) =\diag \left( \begin{bmatrix} \alpha_1 e^{j 2 \pi k T_o f_{D,1}}  & \ldots & \alpha_M e^{j 2 \pi k T_o f_{D,M}}\end{bmatrix} \right)
\)
and per-subcarrier delay responses
\(
\pmb{D}_n(\pmb{\tau}) =\diag \left(\begin{bmatrix} c_n(\tau_0)  & \ldots & c_n(\tau_M) \end{bmatrix}\right)
\).
For \ac{ULA} configurations,
\begin{equation*}
\begin{split}
    \pmb{a}_t(\phi) &= \exp\!\Big(-j \tfrac{2 \pi d_{\tt t}}{\lambda}\, [0, \ldots, N_{\tt t}-1]^{T}\, \sin\phi\Big) \in \mathbb{C}^{N_{\tt t} \times 1}, \\
    \pmb{a}_r(\theta) &= \exp\!\Big(-j \tfrac{2 \pi d_{\tt r}}{\lambda}\, [0, \ldots, N_{\tt r}-1]^{T}\, \sin\theta\Big) \in \mathbb{C}^{N_{\tt r} \times 1},
\end{split}
\end{equation*}
with antenna spacings \(d_{\tt t},d_{\tt r}\) and wavelength \(\lambda\).
The complex coefficients $\alpha_m$ in $\pmb{G}_k(\pmb{f_D})$ are the \ac{CPI}-constant per-path gains defined in~\eqref{eq:alpha_factor_local_paper}. The geometric parameters $(\phi_m,\theta_m,\tau_m)$ are encoded by the spatial and frequency slopes and are unaffected by the scaling in $\alpha_m$. Under calibrated \acp{ULA} with half-wavelength spacing, angles are taken on the principal domain $\phi,\theta\in[-90^\circ,90^\circ]$. We also restrict delays to $\tau\in[0,T_{\tt CP})\subset[0,1/\Delta_f)$, which avoids delay aliasing and guarantees unambiguous \ac{ToA}, consistent with a communication-centric design.

In low-mobility scenarios where Doppler shifts are negligible (i.e., \(f_{D,m} \approx 0\)), the frequency domain \ac{CIR} from~\eqref{eq:CIR-freq-doppler} simplifies to
\begin{equation}
\label{eq:CIR-freq}
    \pmb{H}_n
    =
    \pmb{A}_r(\pmb{\Theta})
    \pmb{G}
    \pmb{D}_n(\pmb{\tau})
    \pmb{A}_t^T(\pmb{\Phi}),
\end{equation}
\noindent where the dependence on $k$ and Doppler shifts are omitted, and $ \pmb{G} =\diag \left( \begin{bmatrix} \alpha_1 & \ldots & \alpha_M \end{bmatrix} \right)$. At higher carrier frequencies or under mobility, the Doppler term in \eqref{eq:CIR-freq-doppler} must be retained.
\subsection{Received Signal Model}
The radar unit, equipped with a \ac{ULA} of \(N_{\tt{r}}\) antennas, receives the downlink \ac{OFDM} symbols transmitted by the \ac{BS}. Combining the transmitted signal in ~\eqref{eq:tx-OFDM-DL} with the channel response in ~\eqref{eq:CIR} and applying \ac{FFT}, the received signal on the \(n^{\text{th}}\) subcarrier and \(k^{\text{th}}\) \ac{OFDM} symbol is given by
\begin{equation}
    \label{eq:sys_model}
    \pmb{y}_{n,k} = \pmb{H}_{n,k} \pmb{s}_{n,k} + \pmb{w}_{n,k}
    \in  \mathbb{C}^{N_r \times 1},  
\end{equation}
\noindent where $\pmb{H}_{n,k}$ is the bistatic radar channel frequency response for the $n^{\text{th}}$ subcarrier and $k^{\text{th}}$ \ac{OFDM} symbol. Here, $\pmb{w}_{n,k} \in \mathbb{C}^{N_{\tt{r}} \times 1}$ is an \ac{AWGN} vector with zero mean and covariance \(\sigma^2 \pmb{I}\).
\subsection{Channel Estimation}
\label{subsec:channel_estimation}
In this work, the communication function operates unchanged, without being disrupted or sacrificed, and the passive radar opportunistically performs sensing by reusing the \ac{BS} downlink waveform. Specifically, the sensing receiver forms the per-subcarrier channel estimates from standard downlink reference signals, and when a backhaul is available it may optionally leverage the decoded payload symbols, though these are not required. This communication-centric design keeps the pilot structure, frame format, scheduling, and precoding unchanged and preserves the full bandwidth for communication, thereby avoiding any trade-off between sensing accuracy and communication requirements. The only incremental cost is local computation at the radar receiver.
Because sensing reuses the communication waveform and its \ac{CP}, the maximum unambiguous excess delay of any bistatic echo is bounded by the \ac{CP}. For target $m$, the \ac{TDoA} (relative to the direct \ac{BS} to radar path) is
\begin{equation}
\label{eq:excess_path}
\tau_m^{(\mathrm{ex})}
\;\triangleq\;
\frac{R_{{\tt t},m}+R_{{\tt r},m}-R_{{\tt t},{\tt r}}}{c}
\;<\; T_{\tt CP},
\end{equation}
where $R_{{\tt t},m}$ and $R_{{\tt r},m}$ are the \ac{BS} to target and target to radar ranges and $R_{{\tt t},{\tt r}}$ is the direct \ac{BS} to radar distance. Equivalently, the excess path length must satisfy $R_{{\tt t},m}{+}R_{{\tt r},m}{-}R_{{\tt t},{\tt r}}<c\,T_{\tt CP}$. Enforcing \eqref{eq:excess_path} ensures that echoes remain confined within the \ac{CP}, preserving \ac{OFDM} orthogonality and avoiding \ac{ISI}/\ac{ICI}. This induces a standard communication-centric trade-off: increasing $T_{\tt CP}$ enlarges the unambiguous excess path window at the cost of higher overhead, whereas reducing $T_{\tt CP}$ improves spectral efficiency but shrinks bistatic coverage. All simulated geometries in this work satisfy \eqref{eq:excess_path}.

The proposed methods rely on accurately estimating the radar sensing channel, which encodes key sensing parameters such as \ac{AoA}, \ac{AoD}, and \ac{ToA}.
We assume the complex path gains and sensing parameters are time-invariant over a \ac{CPI}, which typically lasts a few milliseconds for environments with moderate-speed targets \cite{9540344}.
Let $\pmb{S}_{n} \in \mathbb{C}^{N_{\tt{t}} \times K_{\tt{P}}}$ represent the matrix of $K_{\tt{P}}$ known transmitted \ac{OFDM} symbols on the $n^{\text{th}}$ subcarrier, which are provided to the passive radar via the backhaul connection with the \ac{BS} \cite{bazzi_backhaul}. The $k^{\text{th}}$ column of $\pmb{S}_{n}$ is given by \(\pmb{S}_{n[:,k]} = \begin{bmatrix} \pmb{s}_{1,k} & \pmb{s}_{2,k} & \cdots & \pmb{s}_{N_{\tt{t}}, k} \end{bmatrix}^T\). Similarly, let $\pmb{Y}_{n} \in \mathbb{C}^{N_{\tt{r}} \times K_{\tt{P}}}$ denote the matrix of received symbols $\lbrace \pmb{y}_{n,k} \rbrace_{k=1}^{K_{\tt{P}}}$ on the $n^{\text{th}}$ subcarrier. The objective is to estimate the channel response $\pmb{H}_{n,k}$ using these transmitted and received symbol matrices. 
For the $n^{\text{th}}$ subcarrier, the \ac{LS} estimator provides a straightforward channel estimate as follows
\begin{equation}
    \label{eq:method2_step0}
    \begin{split}
    \bar{\pmb{H}}_n &= \pmb{Y}_{n} \pmb{S}_{n}^H (\pmb{S}_{n} \pmb{S}_{n}^H)^{-1} = \pmb{Y}_{n} \pmb{S}_{n}^\dagger,
    \end{split}
\end{equation}
where $\pmb{S}_{n}^\dagger$ denotes the Moore-Penrose pseudo-inverse of $\pmb{S}_{n}$, given by $\pmb{S}_{n}^\dagger = \pmb{S}_{n}^H (\pmb{S}_{n} \pmb{S}_{n}^{H})^{-1}$ for each subcarrier $n$. 
As is standard in communication receivers, per-subcarrier channel estimates $\bar{\pmb{H}}_{n}$ are normally formed for tasks such as equalization, demodulation, and decoding, and we therefore adopt $\bar{\pmb{H}}_{n}$ as the sensing input in our communication-centric setting.

Unless stated otherwise, the radar is provisioned with the \ac{BS} transmit array configuration (\ac{ULA} with known inter-element spacing and orientation) to form the steering vectors. 
Moreover, we restrict to unprecoded references so that the downlink beamforming is the identity, and we allocate $K_{\tt{P}} \geq N_{\tt{t}}$ symbols to ensure that $\operatorname{rank}(\pmb{S}_n)=N_{\tt t}$ and that the \ac{LS} estimate $\bar{\pmb{H}}_n$ exists.

Consider a sequence of sub-frames indexed by \(p\), where each sub-frame comprises \(\Bar{K}_{\tt{P}} \geq N_{\tt{t}}\) \ac{OFDM} symbols. Specifically, the \(p^{\text{th}}\) sub-frame contains the \ac{OFDM} symbols indexed by \(k = \left( (p-1)\Bar{K}_{\tt{P}} + 1 \right), \ldots, p \Bar{K}_{\tt{P}}\). The \ac{LS} channel estimate for the \(n^{\text{th}}\) subcarrier in the \(p^{\text{th}}\) sub-frame is expressed as
\begin{equation}
\label{eq:estim_sub_frames}
\bar{\pmb{H}}_{n, p} = \pmb{Y}_{n, p} \pmb{S}_{n, p}^\dagger,
\end{equation}
\noindent where \(\pmb{Y}_{n, p}\) and \(\pmb{S}_{n, p}\) denote the matrices of received and transmitted symbols, respectively, for the \(p^{\text{th}}\) sub-frame.

In cases where Doppler shifts are negligible, the estimated channel response on subcarrier \(n\), \(\bar{\pmb{H}}_n\), can be expressed as
\begin{equation}
    \begin{split}
    \label{eq:explanation_H_sum_noise}
    \bar{\pmb{H}}_n &= \pmb{Y}_{n} \pmb{S}_{n}^H (\pmb{S}_{n} \pmb{S}_{n}^H)^{-1} \\
    &= \left(\pmb{H}_n \pmb{S}_{n} + \pmb{W}_{n}  \right) \pmb{S}_{n}^H (\pmb{S}_{n} \pmb{S}_{n}^H)^{-1} \\
    &= \pmb{H}_n \pmb{S}_{n} \pmb{S}_{n}^H (\pmb{S}_{n} \pmb{S}_{n}^H)^{-1}+ \pmb{W}_{n}  \pmb{S}_{n}^H (\pmb{S}_{n} \pmb{S}_{n}^H)^{-1} \\
    &=\pmb{H}_n + \pmb{W}_{n} \pmb{S}_{n}^\dagger,
    \end{split}
\end{equation}
\noindent where $\pmb{W}_{n} \pmb{S}_{n}^\dagger$ represents the noise term induced by the additive noise $\lbrace \pmb{w}_{n,k} \rbrace_{k=1}^{K_{\tt{P}}}$ in the received data.
To compile the \ac{CSI} across all subcarriers, the per-subcarrier estimates \(\bar{\pmb{H}}_n\) are assembled into a single matrix \(\pmb{\bar{H}}\) as follows
\begin{equation}
    \begin{split}
    \label{eq:method2_step0_1}
    \pmb{\bar{H}} &= \begin{bmatrix} \vectorize(\bar{\pmb{H}}_1) & \vectorize(\bar{\pmb{H}}_2) & \cdots & \vectorize(\bar{\pmb{H}}_{N_{\tt{P}}}) \end{bmatrix} \\
    &= \pmb{H} + \pmb{\bar{W}},
    \end{split}
\end{equation}
\noindent where $\pmb{\bar{H}} \in \mathbb{C}^{N_{\tt{t}} N_{\tt{r}} \times N_{\tt{P}}}$ represents the frequency-domain channel estimates, $\pmb{H}$ denotes the true \ac{CSI}, and $\pmb{\bar{W}}$ is the aggregated noise matrix.
While the \ac{LS} estimator is computationally efficient, it is sensitive to noise, particularly in low \ac{SNR} conditions. Alternative approaches, such as the \ac{MMSE} estimator, can improve robustness by incorporating prior knowledge of the channel and noise covariance, albeit with a higher computational cost. Regularized \ac{LS} techniques or hybrid methods may also offer a balanced trade-off between robustness and computational efficiency \cite{SURVEYCHANNELESTIMATION}.

In distributed deployments, a per-radar wired backhaul is not required. The channel estimates can be formed from broadcast pilots, with optional semi-blind refinements using locally received payload~\cite{semi_blind}, and timing/frequency can be synchronized over the air (e.g., via the \ac{LoS} \ac{BS} to radar component)~\cite{multistatic_sync}. This preserves the communication waveform and scales to multiple passive receivers.
\section{Proposed Algorithms}
\label{sec:ALGOS}
In this section, we address the problem of sensing parameter estimation within the bistatic communication-centric \ac{ISAC} framework introduced in Section~\ref{sec:system_model}. Accurate estimation of the sensing parameters is essential for achieving the dual functionalities of communication and opportunistic sensing. However, standard approaches, such as the \ac{MLE}, are often computationally prohibitive due to the extensive multi-dimensional optimization required. To provide context, we first present the \ac{MLE} formulation and discuss its computational limitations. We then introduce two proposed methods, IFFT-\ac{C2VNN} and PARAMING, which are designed to strike an effective balance between computational efficiency and estimation accuracy.
\subsection{Maximum Likelihood Parameter Estimation}
\label{subsec:mle}
The \ac{MLE} is widely regarded for its asymptotic efficiency in joint parameter estimation. Nevertheless, it incurs substantial computational costs as it requires a multi-dimensional search over continuous parameter spaces. To jointly estimate \ac{AoA}, \ac{AoD}, and \ac{ToA} parameters, the \ac{MLE} can be formulated by modeling the observed data as deterministic sequences. Consequently, the joint likelihood function of the observed data \(\pmb{\mathcal{Y}}\), conditioned on known pilot signals \(\pmb{\mathcal{S}}\), noise variance \(\sigma^2\), path gains \(\pmb{\alpha} = [\alpha_0, \ldots, \alpha_M]\), \ac{AoA} values \(\pmb{\Theta}\), \ac{AoD} values \(\pmb{\Phi}\), and \ac{ToA} values \(\pmb{\tau}\), is given by
\begin{equation}
\label{eq:likelihood}
\begin{split}
	f(\pmb{\mathcal{Y}} | \pmb{\mathcal{S}}, & \sigma^2, \pmb{\alpha}, \pmb{\Theta}, \pmb{\Phi}, \pmb{\tau}) = 
	\prod_{n=1}^{N_{\tt{P}}} \prod_{k=1}^{K_{\tt{P}}} 
	\frac{1}{\pi \det(\sigma^2 \pmb{I})} \\
	& \times \exp 
	\left(
	-\frac{1}{\sigma^2}
	\left\Vert 
	\pmb{y}_{n,k}
	-
	\pmb{H}_n(\pmb{\alpha}, \pmb{\Theta}, \pmb{\Phi}, \pmb{\tau})
	\pmb{s}_{n,k}
	\right\Vert^2
	\right),
\end{split}
\end{equation}
\noindent where \(\pmb{\mathcal{Y}}\) is constructed by stacking the received signal vectors \(\pmb{y}_{n,k} \in \mathbb{C}^{N_{\tt{r}} \times 1}\), with its \(k^{\text{th}}\) column given by \(\pmb{\mathcal{Y}}_{[:,k]} = \begin{bmatrix} \pmb{y}_{1,k} & \pmb{y}_{2,k} & \cdots & \pmb{y}_{N_{\tt{P}}, k} \end{bmatrix}^T\). Similarly, \(\pmb{\mathcal{S}}\) represents the known signals transmitted by the \ac{BS}, with each column \(\pmb{\mathcal{S}}_{[:,k]} = \begin{bmatrix} \pmb{s}_{1,k} & \pmb{s}_{2,k} & \cdots & \pmb{s}_{N_{\tt{P}}, k} \end{bmatrix}^T\). Here, \(\pmb{H}_n(\pmb{\alpha}, \pmb{\Theta}, \pmb{\Phi}, \pmb{\tau})\) denotes the radar channel frequency response for the \(n^{\text{th}}\) subcarrier, parameterized by the path gains \(\pmb{\alpha}\), angles \(\pmb{\Theta}\) and \(\pmb{\Phi}\), and delays \(\pmb{\tau}\).
For simplicity, we express the log-likelihood as
\begin{equation*}
\label{eq:log-likelihood}
\begin{split}
	\mathcal{L} &\triangleq \log f(\pmb{\mathcal{Y}})
        =
        g(\sigma^2)
        -
        \frac{1}{\sigma^2}
	\sum\limits_{n=1}^{N_{\tt{P}}}
	\sum\limits_{k=1}^{K_{\tt{P}}}
	\big\Vert 
	\pmb{y}_{n,k}
	-
	\pmb{H}_n
	\pmb{s}_{n,k}
	\big\Vert^2,
\end{split}
\end{equation*}
where \(g\) is a function of the noise variance \(\sigma^2\). Consequently, the \ac{MLE} criterion can be formulated as
\begin{equation}
\label{eq:MLE-criterion}
    \argmin_{\pmb{\alpha},\pmb{\Theta},\pmb{\Phi},\pmb{\tau}}
    \Big\Vert 
    \pmb{\mathcal{Y}}
    -
    [ \pmb{I}_{N_{\tt{P}}} \otimes \pmb{A}_r(\pmb{\Theta})\pmb{G} ]
    \pmb{D}(\pmb{\tau})
    [ \pmb{I}_{N_{\tt{P}}} \otimes \pmb{A}_t^T(\pmb{\Phi}) ]
    \pmb{\mathcal{S}}
    \Big\Vert^2,
\end{equation}
\noindent where \(\pmb{D}(\pmb{\tau}) = \diag\big(\pmb{D}_1(\pmb{\tau}), \ldots, \pmb{D}_{N_{\tt{P}}}(\pmb{\tau}) \big)\) represents the delay matrix encoding the \ac{ToA} information.
This estimation problem can be cast as a nonlinear \ac{LS} optimization over continuous parameter spaces. An exhaustive grid search for \ac{MLE} estimation in \eqref{eq:MLE-criterion} results in a prohibitive computational complexity
\begin{equation}
\label{eq:complexity_MLE}
\mathcal{O}(G_\tau^M G_\theta^M G_\phi^M G_\alpha^{2M} \cdot (N_{\tt{r}} N_{\tt{t}} M^2 N_{\tt{P}}^4 + N_{\tt{r}} N_{\tt{P}}^3 K_{\tt{P}})),
\end{equation}
\noindent where \(G_\tau\), \(G_\theta\), \(G_\phi\), and \(G_\alpha\) denote the grid sizes for \ac{ToA}, \ac{AoA}, \ac{AoD}, and path gains, respectively. Here, the terms \(G_\tau^M\), \(G_\theta^M\), and \(G_\phi^M\) reflect the exponential scaling of complexity with grid points for each parameter across the \(M\) paths, while the factors \(N_{\tt{r}} N_{\tt{t}} M^2 N_{\tt{P}}^4\) and \(N_{\tt{r}} N_{\tt{P}}^3 K_{\tt{P}}\) capture the computation required for each grid point over transmit and receive antennas, subcarriers, and symbols, thereby underscoring the infeasibility of brute-force \ac{MLE} for real-time applications.

While the search space over the sensing parameters is indeed continuous, a common \ac{MLE} approach would be to discretize the search space on a uniform grid, which becomes  unfeasible for higher number of targets and sensing parameters. Alternatives can be employed such as the \ac{SAGE} algorithm. Specifically, the \ac{SAGE} algorithm uses alternating expectation-maximization steps, while Richter’s \ac{MLE} (RiMax) leverages gradient-based techniques to explore the likelihood surface. Subspace methods, such as \ac{MUSIC} and \ac{ESPRIT}, can further reduce the complexity by exploiting the signal and noise subspaces for parameter estimation, although these methods may experience performance degradation in low \ac{SNR} scenarios or when scatterers are closely spaced.

In this work, we propose two novel approaches for joint \ac{AoA}/\ac{AoD}/\ac{ToA} estimation: IFFT-\ac{C2VNN} and PARAMING. IFFT-\ac{C2VNN} leverages complex-valued \acp{CNN} to directly predict the sensing parameters from the estimated \ac{CSI}, providing computational efficiency without compromising estimation accuracy. On the other hand, PARAMING exploits the structured characteristics of the system model, incorporating \ac{CSI} estimates to achieve an efficient and accurate solution. Both methods are thus designed to deliver high estimation accuracy, while significantly reducing the computational complexity compared to \ac{MLE}.
\subsection{Deep Learning-based Estimation: IFFT-C2VNN}
\label{sec:ML_method}
The IFFT-\ac{C2VNN} algorithm is a \ac{DL} architecture tailored for estimating the sensing parameters efficiently from the estimated \ac{CSI} matrix $\pmb{\bar{H}}$. In fact, complex-valued \acp{CNN} have demonstrated effectiveness in handling multidimensional data \cite{trabelsi2018deep}, making them ideal for radar and signal processing tasks \cite{cnn_direction_arrival, surveycvnn}. By leveraging the \ac{CSI} structure, IFFT-\ac{C2VNN} provides accurate estimates while addressing the inefficiencies inherent in \ac{MLE} and subspace-based methods.
\begin{figure*}[!ht]
	\centering
    \includegraphics[trim={45mm 116mm 4mm 110mm},clip,width=0.92\textwidth]{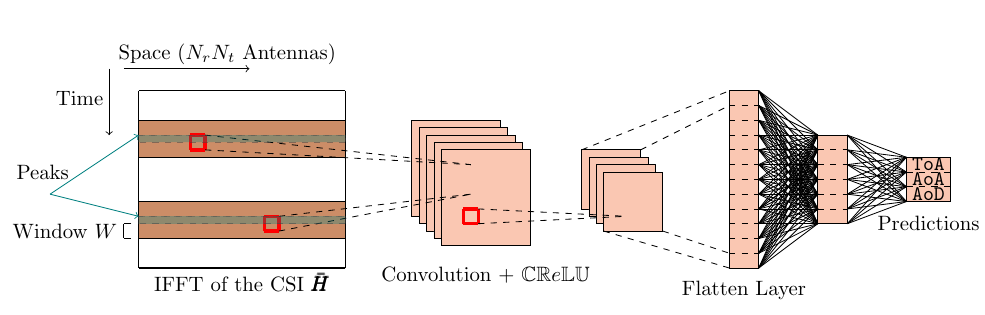}
	\caption{Architecture of the IFFT-\ac{C2VNN} algorithm, consisting of complex-valued convolutional layers and $\mathbb{C}\operatorname{ReLU}$ activation functions.}
	\label{fig_cnn_architecture}
\end{figure*}
\subsubsection{Input Processing}
Given the estimated \ac{CSI} matrix \(\pmb{\bar{H}}\), as defined in \eqref{eq:method2_step0_1}, we can decompose it in terms of the sensing parameters as follows
\begin{equation}
\label{method2_eq:system-model}
\pmb{\bar{H}} = \pmb{B}(\pmb{\Theta}, \pmb{\Phi}) \pmb{G} \pmb{C}^T(\pmb{\tau}) + \widetilde{\pmb{W}},
\end{equation}
where \(\pmb{B}(\pmb{\Theta}, \pmb{\Phi})\) is the spatial response matrix, defined as
\begin{equation*}
\pmb{B}(\pmb{\Theta}, \pmb{\Phi}) = \begin{bmatrix} \pmb{a}_t(\phi_0) \otimes \pmb{a}_r(\theta_0) & \cdots & \pmb{a}_t(\phi_M) \otimes \pmb{a}_r(\theta_M) \end{bmatrix},
\end{equation*}
and the matrix \(\pmb{C}(\pmb{\tau}) = \begin{bmatrix} \pmb{c}(\tau_0) & \cdots & \pmb{c}(\tau_M) \end{bmatrix} \in \mathbb{C}^{N_{\tt{P}} \times (M+1)}\) encapsulates the \ac{ToA} information across the subcarriers, while \(\widetilde{\pmb{W}}\) represents the overall noise term aggregated across subcarriers and transmit-receive antenna pairs.
To process \(\pmb{\bar{H}}\) as input, we first apply an \ac{IDFT} across the subcarrier axis. Let \( \pmb{F} \in \mathbb{C}^{N_{\tt{P}} \times N_{\tt{P}}}\) denote the \ac{DFT} matrix, with each element of \(\pmb{F}\) defined as
\begin{equation}
\left[\pmb{F}\right]_{n,k} = \frac{1}{\sqrt{N_{\tt{P}}}} e^{-j \frac{2 \pi}{N_{\tt{P}}} nk}, \quad n, k = 0, 1, \ldots, N_{\tt{P}}-1.
\end{equation}
The transformation applied to the transpose of \(\pmb{\bar{H}}\) yields
\begin{equation}
\pmb{F}^H \pmb{\bar{H}}^T = \pmb{F}^H \pmb{C}(\pmb{\tau}) \pmb{G} \pmb{B}^T(\pmb{\Theta}, \pmb{\Phi}) + \pmb{F}^H \widetilde{\pmb{W}}.
\end{equation}

To further examine \(\pmb{F}^H \pmb{C}(\pmb{\tau})\), we apply an \ac{IDFT} to each column of the delay matrix \(\pmb{C}(\pmb{\tau})\). The \((p, m)\)-th entry of the resulting matrix \(\pmb{F}^H \pmb{C}(\pmb{\tau})\) is given by
\begin{equation}
\begin{split}
\left[\pmb{F}^H \pmb{C}(\pmb{\tau})\right]_{p,m} &= \frac{1}{\sqrt{N_{\tt{P}}}} \sum_{n=0}^{N_{\tt{P}}-1} e^{j \frac{2 \pi}{N_{\tt{P}}} p n} e^{-j 2 \pi n \Delta_f \tau_m} \\
& = \frac{1}{\sqrt{N_{\tt{P}}}} \sum_{n=0}^{N_{\tt{P}}-1} e^{j 2 \pi n \left( \frac{p}{N_{\tt{P}}} - \Delta_f \tau_m \right)}.
\end{split}
\end{equation}

For non-fractional delays, where each delay \(\tau_m\) aligns with an integer multiple of the sampling interval \(\Delta_{\tt{t}} = \frac{1}{N_{\tt{P}} \Delta_f}\), the term \(\frac{p}{N_{\tt{P}}} - \Delta_f \tau_m\) becomes an integer for certain indices \(p\). In this case, if \(\tau_m = \frac{k}{N_{\tt{P}} \Delta_f}\) for an integer \(k\), we have
\begin{equation}
\left[\pmb{F}^H \pmb{C}(\pmb{\tau})\right]_{p,m} = \begin{cases}
\sqrt{N_{\tt{P}}}, & p = k, \\
0, & \text{otherwise}.
\end{cases}
\end{equation}

The resulting sparsity in \(\pmb{F}^H \pmb{C}(\pmb{\tau})\), with non-zero entries only at rows corresponding to each discrete delay \(\tau_m\), allows each target's delay to occupy a distinct row in \(\pmb{F}^H \pmb{C}(\pmb{\tau})\), facilitating a straightforward extraction of the individual target parameters. This structured sparsity is preserved in \(\pmb{F}^H \pmb{\bar{H}}^T\) under non-fractional delay assumptions, where each delay \(\tau_m\) corresponds to a unique row index \(\widehat{i}_m = N_{\tt{P}} \Delta_f \tau_m\). Consequently, \(\pmb{F}^H \pmb{\bar{H}}^T\) exhibits non-zero entries only in rows indexed by \(\{\widehat{i}_1, \ldots, \widehat{i}_M\}\), with other rows containing near-zero values. This structured sparsity greatly simplifies the parameter estimation by isolating each target's delay contribution, thus supporting efficient extraction of individual sensing parameters.

In the general case with fractional delays, where \(\tau_m\) is not an integer multiple of the sampling interval \(\Delta_{\tt{t}} \), \(\frac{p}{N_{\tt{P}}} - \Delta_f \tau_m\) is generally non-integer. This causes each delay's contribution to spread across multiple rows in \(\pmb{F}^H \pmb{C}(\pmb{\tau})\) and produces a less sparse structure in \(\pmb{F}^H \pmb{\bar{H}}^T\). Here, each delay's influence is represented by a sinc-like spread, with energy dispersed over multiple rows, generating phase variations across a range of rows. To manage this, we apply a windowed input representation centered around the peak row for each target's primary delay index, forming an input tensor \(\bar{\pmb{\mathcal{I}}}_{m} \in \mathbb{C}^{(2W+1) \times N_{\tt{r}} N_{\tt{t}}}\) for the $m^{\text{th}}$ target, as defined by
\begin{equation}
\label{input_windowed}
\bar{\pmb{\mathcal{I}}}_{m}
=
\begin{bmatrix}
\pmb{\bar{{h}}}_{[\widehat{i}_m - W], 1} & \cdots & \pmb{\bar{{h}}}_{[\widehat{i}_m - W], N_{\tt{r}} N_{\tt{t}}} \\
\vdots & \ddots & \vdots \\
\pmb{\bar{{h}}}_{[\widehat{i}_m], 1} & \cdots & \pmb{\bar{{h}}}_{[\widehat{i}_m], N_{\tt{r}} N_{\tt{t}}} \\
\vdots & \ddots & \vdots \\
\pmb{\bar{{h}}}_{[\widehat{i}_m + W], 1} & \cdots & \pmb{\bar{{h}}}_{[\widehat{i}_m + W], N_{\tt{r}} N_{\tt{t}}}
\end{bmatrix},
\end{equation}
where \(\pmb{\bar{{h}}}_{[p], j}\) denotes the $p^{\text{th}}$ row and $j^{\text{th}}$ column entry in \(\pmb{F}^H \pmb{\bar{H}}^T\). The window of size \((2W+1)\) effectively isolates each target’s delay contributions, enabling the \ac{NN} to capture fractional delay information from the phase variations in the local neighborhood and accurately estimate the sensing parameters.
Moreover, the complex gains can be recovered by an \ac{LS} projection onto the recovered manifold
\begin{equation}
\label{eq:alpha_LS}
\hat{\boldsymbol\alpha}=(\Psi^{\!H}\Psi)^{-1}\Psi^{\!H}\pmb{y},
\quad
\pmb{y}=\mathrm{vec}\!\left([\bar{\pmb{H}}_1,\ldots,\bar{\pmb{H}}_{N_{\tt P}}]\right),
\end{equation}
where the $m^{\text{th}}$ column of $\Psi$ is $\mathrm{vec}\!\big(\pmb{a}_{\tt r}(\hat\theta_m)\,\pmb{a}_{\tt t}^{T}(\hat\phi_m)\big)\otimes \pmb{c}(\hat\tau_m)$. This step ensures a scale-consistent $\hat{\boldsymbol\alpha}$ aligned with the predicted geometry. Optionally, the network can be extended to regress the amplitudes and gains of \(\alpha_m\) directly.

It is crucial to note that the current approach assumes resolvable scatterers, where each peak corresponds to a distinct scatterer. However, in cases of non-resolvability, where multiple scatterers contribute to a single peak, a preliminary classification step is required to estimate the number of paths within each peak. In such cases, non-resolvability is typically mitigated in the angular domain, as it is highly improbable for two targets to be sufficiently close in space to produce identical \ac{AoA}, \ac{AoD}, and delay combinations. This classification guarantees that the fixed-size \ac{NN} output aligns with the expected number of parameters to estimate, thereby ensuring a reliable sensing performance.
\subsubsection{Network Architecture}
The IFFT-\ac{C2VNN} architecture, illustrated in Fig.~\ref{fig_cnn_architecture}, is designed to process the windowed input representations $\bar{\pmb{\mathcal{I}}}_{m}$ for each target, facilitating the efficient estimation of the \ac{ToA}, \ac{AoA}, and \ac{AoD} parameters. The network structure includes two complex-valued convolutional layers, each followed by a $\mathbb{C}\operatorname{ReLU}$ activation function, and concludes with two complex linear layers that output the estimated sensing parameters.
Each convolutional layer comprises $10$ complex filters, with the $n^{\text{th}}$ filter defined as \(\pmb{W}_n = \pmb{W}_n^{r} + j \pmb{W}_n^{i}\).
For a complex input \(\pmb{z} = \pmb{x} + j \pmb{y}\), where \(\pmb{x}\) and \(\pmb{y}\) denote the real and imaginary parts, respectively, the resulting feature map \(\pmb{\widehat{z}}\) from the convolutional layer is computed as
\begin{equation}
\pmb{\widehat{z}} = \left(\pmb{W}_n^r * \pmb{x} - \pmb{W}_n^i * \pmb{y}\right) + j\left(\pmb{W}_n^r * \pmb{y} + \pmb{W}_n^i * \pmb{x}\right).
\end{equation}
The complex-valued linear layers follow a similar structure, with weights and biases represented as \(\pmb{W} = \pmb{W}^r + j \pmb{W}^i\) and \(\pmb{b} = \pmb{b}^r + j \pmb{b}^i\), respectively. Given an input \(\pmb{z} = \pmb{x} + j \pmb{y}\), the output \(\pmb{\widehat{z}}\) of the linear layer is calculated as
\begin{equation}
\pmb{\widehat{z}} = \left(\pmb{W}^r \pmb{x} - \pmb{W}^i \pmb{y} + \pmb{b}^r\right) + j\left(\pmb{W}^r \pmb{y} + \pmb{W}^i \pmb{x} + \pmb{b}^i\right).
\end{equation}
\noindent The $\mathbb{C}\operatorname{ReLU}$ activation function introduces non-linearity, enabling the network to effectively model complex-domain non-linear functions. This function is defined as
\begin{equation}
\mathbb{C}\operatorname{ReLU}(\pmb{\widehat{z}})=\operatorname{ReLU}(\Re(\pmb{\widehat{z}}))+j \operatorname{ReLU}(\Im(\pmb{\widehat{z}})), 
\end{equation}
where $\operatorname{ReLU}(x) = \max(0,x)$.
\subsubsection{Training Procedure and Hyperparameters}
The network is trained using a synthetic dataset comprising \( E \) simulations, i.e. training points, generated in accordance with the system model outlined in Section~\ref{sec:system_model}. In each simulation, \( M \) targets are randomly positioned within the environment. The training objective is to minimize the \ac{MSE} between the predicted and true parameters, as defined by
\begin{equation}
\begin{split}
L_{\text{MSE}} = \frac{1}{M E} \sum_{e=1}^E & \sum_{m=1}^M \bigg[ \left(\widehat{\theta}_m^{\operatorname{tar}}(e)-\theta_{m}(e)\right)^2 + \\
& \left(\widehat{\phi}_m^{\operatorname{tar}}(e)-\phi_{m}(e)\right)^2 + \left(\widehat{\tau}_m^{\operatorname{tar}}(e)-\tau_{m}(e)\right)^2 \bigg],
\end{split}
\end{equation}
\noindent where \( \widehat{\theta}_m^{\operatorname{tar}}(e) \), \( \widehat{\phi}_m^{\operatorname{tar}}(e) \), and \( \widehat{\tau}_m^{\operatorname{tar}}(e) \) represent the predicted \ac{AoA}, \ac{AoD}, and \ac{ToA} for the \( m^{\text{th}} \) target in the \( e^{\text{th}} \) training point, respectively.
\subsection{PARAMING}
\label{sec:3d_algorithm}
The PARAMING algorithm offers an efficient method for extracting the joint \ac{AoA}, \ac{AoD}, and \ac{ToA} parameters from the \ac{CSI} matrix \(\pmb{\bar{H}}\). In contrast to traditional \ac{MLE} or subspace-based methods, which rely on exhaustive grid searches or complex multi-dimensional optimizations, PARAMING capitalizes on a structured decomposition of \(\pmb{\bar{H}}\) into sub-array and subcarrier blocks, allowing for an efficient parameter extraction through matrix transformations. Below, we present the matrix construction, parameterization, and algorithmic steps for isolating the sensing parameters.

Given an estimate of the \ac{CSI} \(\pmb{\bar{{H}}}\), PARAMING introduces sub-array sizes \(1 < M_{\tt{t}} \leq N_{\tt{t}}\) and \(1 < M_{\tt{r}} \leq N_{\tt{r}}\) along with a sub-\ac{OFDM} symbol size \(1 < M_{\tt{P}} \leq N_{\tt{P}}\) subcarriers. Then, by leveraging the Vandermonde structure of the transmit/receive \ac{ULA} configurations and the regular subcarrier spacing, a Hankel-block-Hankel-block-Hankel type matrix can be formed using the entries of \(\pmb{\bar{\mathcal{H}}}\) as follows
\begin{equation}
\label{method2_step1-1}
\pmb{\bar{\mathcal{H}}} = \begin{bmatrix}
		\bar{\pmb{\mathcal{H}}}_1 & \bar{\pmb{\mathcal{H}}}_2 & \dots & \bar{\pmb{\mathcal{H}}}_{K_{\tt{t}}} \\
		\bar{\pmb{\mathcal{H}}}_2 & \bar{\pmb{\mathcal{H}}}_3 & \dots & \bar{\pmb{\mathcal{H}}}_{K_{\tt{t}} + 1} \\
		\vdots & \vdots & \ddots & \vdots \\
		\bar{\pmb{\mathcal{H}}}_{M_{\tt{t}}} & \bar{\pmb{\mathcal{H}}}_{M_{\tt{t}} + 1} & \dots & \bar{\pmb{\mathcal{H}}}_{N_{\tt{t}}}
	\end{bmatrix},
\end{equation}
where $K_{\tt{t}} \triangleq N_{\tt{t}} - M_{\tt{t}} + 1$ denotes the number of transmit sub-arrays. Each block \(\bar{\pmb{\mathcal{H}}}_i\) itself has a Hankel-block structure given by
\begin{equation}
\label{method2_step1-2}
\bar{\pmb{\mathcal{H}}}_i = \begin{bmatrix}
		\bar{\pmb{\mathcal{H}}}_{i,1} & \bar{\pmb{\mathcal{H}}}_{i,2} & \dots & \bar{\pmb{\mathcal{H}}}_{i,K_N} \\
		\bar{\pmb{\mathcal{H}}}_{i,2} & \bar{\pmb{\mathcal{H}}}_{i,3} & \dots & \bar{\pmb{\mathcal{H}}}_{i,K_N + 1} \\
		\vdots & \vdots & \ddots & \vdots \\
		\bar{\pmb{\mathcal{H}}}_{i,M_{\tt{P}}} & \bar{\pmb{\mathcal{H}}}_{i,M_{\tt{P}} + 1} & \dots & \bar{\pmb{\mathcal{H}}}_{i,N_{\tt{P}}}
	\end{bmatrix},
\end{equation}
where \(K_N \triangleq N_{\tt{P}} - M_{\tt{P}} + 1\) represents the number of sub-\ac{OFDM} frames, and each entry \(\bar{\pmb{\mathcal{H}}}_{i,j}\) is itself a Hankel matrix formed from the \ac{CSI} entries as
\begin{equation}
\label{method2_step1-3}
\bar{\pmb{\mathcal{H}}}_{i,j} = \begin{bmatrix}
		\bar{h}_{i,j,1} & \bar{h}_{i,j,2} & \dots & \bar{h}_{i,j,K_{\tt{r}}} \\
		\bar{h}_{i,j,2} & \bar{h}_{i,j,3} & \dots & \bar{h}_{i,j,K_{\tt{r}} + 1} \\
		\vdots & \vdots & \ddots & \vdots \\
		\bar{h}_{i,j,M_{\tt{r}}} & \bar{h}_{i,j,M_{\tt{r}} + 1} & \dots & \bar{h}_{i,j,N_{\tt{r}}}
	\end{bmatrix},
\end{equation}
where \(K_{\tt{r}} \triangleq N_{\tt{r}} - M_{\tt{r}} + 1\) is the number of receive sub-arrays. Each element \(\bar{h}_{i,j,k}\) corresponds to an entry in \(\pmb{\bar{H}}\) as defined in \eqref{eq:method2_step0_1}, with a compact indexing notation \(\bar{h}_{i,j,k} = \pmb{\bar{H}}_{[k + (i - 1) N_{\tt{r}}, j]}\).
Using this structured representation, we express \(\pmb{\bar{\mathcal{H}}}\) in terms of the \ac{ToA}, \ac{AoA}, and \ac{AoD} parameters
\begin{equation}
\label{eq:method2_system-model-2}
\pmb{\bar{\mathcal{H}}} = \pmb{\mathcal{B}}_{M_{\tt{r}},M_{\tt{P}},M_{\tt{t}}}(\pmb{\Theta},\pmb{\Phi},\pmb{\tau}) \pmb{G} \pmb{\mathcal{B}}_{K_{\tt{r}},K_N,K_{\tt{t}}}^T(\pmb{\Theta},\pmb{\Phi},\pmb{\tau}) + \widetilde{\pmb{\mathcal{W}}},
\end{equation}
where \(\pmb{G}\) is the path gain matrix, \(\widetilde{\pmb{\mathcal{W}}}\) represents noise, while the manifold \(\pmb{\mathcal{B}}_{n,m,p}(\pmb{\Theta},\pmb{\Phi},\pmb{\tau})\) is defined as
\begin{equation}
\pmb{\mathcal{B}}_{n,m,p}(\pmb{\Theta},\pmb{\Phi},\pmb{\tau}) = \begin{bmatrix}
		\pmb{A}_r(\pmb{\Theta})_{[1:n,:]} \\
		\vdots \\
		\pmb{A}_r(\pmb{\Theta})_{[1:n,:]} \pmb{D}^{m-1}_\tau(\pmb{\tau}) \\
		\pmb{A}_r(\pmb{\Theta})_{[1:n,:]} \pmb{D}_\tau(\pmb{\tau}) \pmb{D}_\phi(\pmb{\Phi}) \\
		\vdots \\
		\pmb{A}_r(\pmb{\Theta})_{[1:n,:]} \pmb{D}^{m-1}_\tau(\pmb{\tau}) \pmb{D}_\phi(\pmb{\Phi}) \\
		\vdots \\
		\pmb{A}_r(\pmb{\Theta})_{[1:n,:]} \pmb{D}_\tau(\pmb{\tau}) \pmb{D}^{p-1}_\phi(\pmb{\Phi}) \\
		\vdots \\
		\pmb{A}_r(\pmb{\Theta})_{[1:n,:]} \pmb{D}^{m-1}_\tau(\pmb{\tau}) \pmb{D}_\phi^{p-1}(\pmb{\Phi})
	\end{bmatrix},
\end{equation}
where $	\pmb{D}_\tau(\pmb{\tau}) =\diag (\begin{bmatrix} \pmb{c}(\tau_1)  & \ldots & \pmb{c}(\tau_M) \end{bmatrix})$ and $	\pmb{D}_\phi(\pmb{\Phi}) =\diag (\begin{bmatrix} \pmb{a}(\phi_1)  & \ldots & \pmb{a}(\phi_M) \end{bmatrix})$.
Then, we control the sub-array sizes to inflate the left/right manifold \(\pmb{\mathcal{B}}_{M_{\tt{r}},M_{\tt{P}},M_{\tt{t}}}\) containing the \ac{ToA}, \ac{AoA} and \ac{AoD} information. To exploit this, we form two interconnected matrices from $\pmb{\bar{\mathcal{H}}}$ as follows
\begin{equation}
	\label{eq:method2_step2_1}
\begin{split}
	\pmb{\bar{\mathcal{H}}}^{(1)} 
	\triangleq
	\pmb{\bar{\mathcal{H}}}_{[:,\mathcal{S}^{(1)}]} = 
	\pmb{\mathcal{B}}_{M_{\tt{r}},M_{\tt{P}},M_{\tt{t}}}(\pmb{\Theta},\pmb{\Phi},\pmb{\tau})
	\pmb{G}
	\pmb{\Pi}^T
	+
	\bar{\pmb{\mathcal{W}}}^{(1)},
\end{split}
\end{equation}
and
\begin{equation}
	\label{eq:method2_step2_2}
\begin{split}
	\pmb{\bar{\mathcal{H}}}^{(2)} 
	\triangleq
	\pmb{\bar{\mathcal{H}}}_{[:,\mathcal{S}^{(2)}]} = 
	\pmb{\mathcal{B}}_{M_{\tt{r}},M_{\tt{P}},M_{\tt{t}}}(\pmb{\Theta},\pmb{\Phi},\pmb{\tau})
	\pmb{G}
	\pmb{D}_\tau(\pmb{\tau})
	\pmb{\Pi}^T
	+
	\bar{\pmb{\mathcal{W}}}^{(2)},
\end{split}
\end{equation}
where $\mathcal{S}^{(1)} = \bigcup\nolimits_{k = 0}^{K_{\tt{t}}-1}  \lbrace \mathcal{S} + k K_{\tt{r}} K_{N_{\tt{P}}}  \rbrace$ and $\mathcal{S}^{(2)} = \mathcal{S}^{(1)} + K_{\tt{r}}$ are sets of selected column indices from \(\pmb{\bar{\mathcal{H}}}\), given $\mathcal{S} = \lbrace 1 , 2 \ldots K_{\tt{r}}(K_{N_{\tt{P}}} - 1) \rbrace$. Here, $\pmb{\Pi} = [\pmb{\mathcal{B}}_{K_{\tt{r}},K_N,K_{\tt{t}}}]_{[\mathcal{S}^{(1)},:]}$, while $\bar{\pmb{\mathcal{W}}}^{(1)}$ and $\bar{\pmb{\mathcal{W}}}^{(2)}$ are noise matrices with elements from from $\widetilde{\pmb{\mathcal{W}}}$.
Subsequently, the matrices \(\pmb{\bar{\mathcal{H}}}^{(1)}\) and \(\pmb{\bar{\mathcal{H}}}^{(2)}\) are exploited to estimate the \acp{ToA} via the combination
\begin{equation}
\begin{split}
	\pmb{\bar{\mathcal{H}}}_{\gamma}
	&\triangleq
	\pmb{\bar{\mathcal{H}}}^{(2)}  - \gamma \pmb{\bar{\mathcal{H}}}^{(1)} \\ &
	=
	\pmb{\mathcal{B}}_{M_{\tt{r}},M_{\tt{P}},M_{\tt{t}}}(\pmb{\Theta},\pmb{\Phi},\pmb{\tau})
	\pmb{G}
	\Big( 
	\pmb{D}_\tau(\pmb{\tau})
	-
	\gamma
	\pmb{I}
	\Big)
	\pmb{\Pi}^T
	+
	\bar{\pmb{\mathcal{W}}}_{\gamma},
\end{split}
\end{equation}
where \(\pmb{\bar{\mathcal{W}}}_{\gamma} = \bar{\pmb{\mathcal{W}}}^{(2)} - \gamma \bar{\pmb{\mathcal{W}}}^{(1)}\). Given that \(\pmb{\Pi}\) and \(\pmb{\mathcal{B}}_{M_{\tt{r}},M_{\tt{P}},M_{\tt{t}}}(\pmb{\Theta},\pmb{\Phi},\pmb{\tau})\) are full-rank matrices, the rank of \(\pmb{\bar{\mathcal{H}}}_{\gamma}\) decreases at \(\gamma = \pmb{c}_1(\tau_m)\) for each \(m\). Thus, we can estimate the \acp{ToA} by performing a grid search over $\tau$ and testing the $M^{\text{th}}$ largest singular value of $\pmb{\bar{\mathcal{H}}}_{\gamma}$, then we choose the $M$ minima of the resulting spectrum. Alternatively, since the aforementioned procedure is highly complex, we resort to a truncated \ac{SVD} method. In essence, we first perform a classical \ac{SVD} on $\pmb{\bar{\mathcal{H}}}^{(1)}$, namely
$
\pmb{\bar{\mathcal{H}}}^{(1)}  = \pmb{U}\pmb{\Sigma}\pmb{V}^H,
$
where $\pmb{U}$ and $\pmb{V}$ represent the left and right singular vectors of $\pmb{\bar{\mathcal{H}}}^{(1)}$, respectively. Moreover, $\pmb{\Sigma}$ is composed of the singular values of $\widehat{\pmb{\mathcal{H}}}^{(1)}$ along its diagonal in decreasing order. After this \ac{SVD}, we perform a truncation by first obtaining a diagonal $M \times M$ matrix $\pmb{\bar{\Sigma}}$ consisting of the strongest $M$ singular values in $\pmb{\Sigma}$. The corresponding left and right singular vectors are stacked in $\pmb{\bar{U}}$ and $\pmb{\bar{V}}$, respectively. Then, we compute 
\begin{equation}
\label{eq:method2_step4}
	\pmb{T} = 
	\pmb{\bar{\Sigma}}^{-1}
	\pmb{\bar{U}}^H
	\pmb{\bar{\mathcal{H}}}^{(2)} 
	\pmb{\bar{V}},
\end{equation}
whose eigenvalues are represented by $\gamma_1 \ldots \gamma_M$. Note that these eigenvalues are estimates of $\pmb{c}_1(\hat{\tau}_m)$. Hence, a simple approach to compute $\lbrace \hat{\tau}_m \rbrace_{m=1}^M$ is via   
\begin{equation}
\label{eq:method2_step6}
	\hat{\tau}_m = - \frac{ \angle \gamma_m }{2 \pi \Delta_f}  , \qquad \forall m = 1 \ldots M.
\end{equation}
\noindent To obtain the \acp{AoD}/\acp{AoA} that are associated with each of the estimated \acp{ToA}. We perform an \ac{LS} fit based on the received signal
\begin{equation}
	\label{eq:method2_step7}
	\widehat{\pmb{Y}}
	=
	\argmin_{\pmb{Y}}
	\big
	\Vert
	\bar{\pmb{H}} 
	-
	\pmb{Y}	
	\pmb{C}^T(\widehat{\pmb{\tau}}) \big\Vert^2
	=
	\bar{\pmb{H}}
	\pmb{C}^*(\widehat{\pmb{\tau}})
	\big(
	\pmb{C}^T(\widehat{\pmb{\tau}})
	\pmb{C}^*(\widehat{\pmb{\tau}})
	\big)^{-1}
	 .
\end{equation}
Note that $\widehat{\pmb{Y}}$ is indeed an \ac{LS} estimate of $\pmb{B}(\pmb{\Theta},\pmb{\Phi})\pmb{G}$, given the \ac{ToA}s. Based on this, we perform a second \ac{LS} stage tailored for estimating the $\pmb{B}(\pmb{\Theta},\pmb{\Phi})$ as such
\begin{equation}
\label{eq:opt-to-solve-B}
\lbrace \widehat{\pmb{B}},\widehat{\pmb{\alpha}} \rbrace
=
\begin{cases}
\argmin_{\pmb{B},\pmb{\alpha}}
 &
\big
\Vert
	\widehat{\pmb{Y}}
	-
	\pmb{B}
	\pmb{G}	
\big\Vert^2, \\
\subjectto &\Vert {\pmb{B}}_{[:,m]}\Vert = 1, \ \pmb{G} = \diag(\pmb{\alpha}).
\end{cases}
\end{equation}
Then, we separate the optimization problem in \eqref{eq:opt-to-solve-B} into $M$ independent problems due to the diagonal structure of $\pmb{G}$, i.e.
\begin{equation}
\widehat{\pmb{B}}_{[:,m]}
=
\begin{cases}
\argmin_{\pmb{b}_m} &
\big
	\Vert
	\widehat{\pmb{Y}}_{[:,m]}
	-
	\alpha_m
	\pmb{b}_m	
\big\Vert^2, \\
\subjectto &\Vert {\pmb{b}_m} \Vert = 1,
\end{cases}
\end{equation}
\noindent where the solution is trivial and is given as
\begin{equation}
\label{eq:method2_step8}
	\widehat{\pmb{b}}_{m} = \frac{\widehat{\pmb{Y}}_{[:,m]}}{\Vert \widehat{\pmb{Y}}_{[:,m]} \Vert}, \qquad \forall m = 1 \ldots M.
\end{equation}
\noindent Moreover, the same column provides a closed-form amplitude estimate
\begin{equation}
\big|\widehat{\alpha}_m\big| \;=\; \frac{\big\Vert \widehat{\pmb{Y}}_{[:,m]} \big\Vert}{\sqrt{N_{\tt t}\,N_{\tt r}}}.
\end{equation}
\noindent Now, given $\{\widehat{\pmb{b}}_{m}\}_{m=1}^M$, we then estimate the \ac{AoA} and \ac{AoD} values via a $2\mathrm{D}$ regression on the phases via the following problem
\begin{equation}
\begin{split}
\label{eq:method2_step09}
(\hat{\theta}_m,\hat{\phi}_m,\hat{\delta}_m)
&=
\argmin_{\theta_m,\phi_m,\delta_m}
\Big
	\Vert
	\angle \widehat{\pmb{b}}_{m}
	-
	\pmb{\Xi}
	\begin{bmatrix}
            \theta_m \\
		\phi_m \\
		\delta_m
	\end{bmatrix}
\Big
	\Vert^2
 =\pmb{\Xi}^\dagger  \angle \widehat{\pmb{b}}_{m} , 
\end{split}
\end{equation}
for all $m$, where $\pmb{\Xi}
	=
	\begin{bmatrix}
		\pmb{1}_{N_{\tt{t}}} \otimes \pmb{e}_{N_{\tt{r}}}, & 
             \pmb{e}_{N_{\tt{t}}} \otimes \pmb{1}_{N_{\tt{r}}},  &
            \pmb{1}_{N_{\tt{t}}N_{\tt{r}}}
	\end{bmatrix} \in \mathbb{C}^{N_{\tt{t}}N_{\tt{r}} \times 3}$,    
and the fitted offset $\hat{\delta}_m$ captures the per-path phase embedded in $\alpha_m$. Combining it with the amplitude above yields the complex gain estimate
\(
\widehat{\alpha}_m \;=\; \big|\widehat{\alpha}_m\big|\, e^{j\hat{\delta}_m},
\)
which can incorporate the bistatic \ac{RCS}, range loss, array/beam gains, and carrier-phase terms, among others.
Moreover, in order to avoid abrupt phase changes, the phases of $\angle \widehat{\pmb{b}}_{m}$ are unwrapped prior to the regression. A summary of the proposed sensing method is given in \textbf{Algorithm~\ref{alg:alg2}}.
\begin{algorithm}[!ht]
\caption{PARAMING (joint \ac{ToA}/\ac{AoA}/\ac{AoD} Estimation)}\label{alg:alg2}
\begin{algorithmic}[1]
\STATE {\textsc{input}: $\lbrace \pmb{Y}_{n}  \rbrace_{n=1}^{N_{\tt{P}}}$, $\lbrace \pmb{S}_{n} \rbrace_{n=1}^{N_{\tt{P}}}$} \\
 {\textsc{Channel Estimation}:} 
\STATE \hspace{0.5cm} Obtain $\lbrace \widehat{\pmb{H}}_n \rbrace_{n=1}^{N_{\tt{P}}}$ according to equation \eqref{eq:method2_step0}\\
 {\textsc{Sensing Estimation}:} 
\STATE\hspace{0.5cm} Arrange ${\pmb{\bar{\mathcal{H}}}}$ as given by equations \eqref{method2_step1-1} and \eqref{method2_step1-2}.
\STATE\hspace{0.5cm} Compute $\pmb{\bar{\mathcal{H}}}^{(1)} , \pmb{\bar{\mathcal{H}}}^{(2)} $ as \eqref{eq:method2_step2_1} and \eqref{eq:method2_step2_2}, respectively.
\STATE\hspace{0.5cm} Get a truncated \ac{SVD} of $\pmb{\bar{\mathcal{H}}}^{(1)} $.
	\begin{equation*}
		\big[ \pmb{\bar{U}}, \pmb{\bar{\Sigma}}, \pmb{\bar{V}}\big] \gets \TSVD_M(\pmb{\bar{\mathcal{H}}}^{(1)} ).
	\end{equation*}
\STATE\hspace{0.5cm} Calculate $\pmb{T}$ via \eqref{eq:method2_step4}, given $\pmb{\bar{U}}, \pmb{\bar{\Sigma}}, \pmb{\bar{V}},\pmb{\bar{\mathcal{H}}}^{(1)} $. 
\STATE\hspace{0.5cm} Get the eigenvalues of $\pmb{T}$, i.e. $\lbrace {\sigma}_m \rbrace_{m=1}^M$.
\STATE\hspace{0.5cm} Estimate $\hat{\tau}_m$ given ${\sigma}_m$ via \eqref{eq:method2_step6}. Repeat for $m = 1 \ldots M$.
\STATE\hspace{0.5cm} Obtain $\widehat{\pmb{Y}}$ through equation \eqref{eq:method2_step7}.
\STATE\hspace{0.5cm} Given $\widehat{\pmb{Y}}$, obtain $\lbrace \widehat{\pmb{b}}_m \rbrace_{m=1}^M$ via \eqref{eq:method2_step8}.
\STATE\hspace{0.5cm} For each $m$, obtain $(\hat{\theta}_m,\hat{\phi}_m)$ as discussed in \eqref{eq:method2_step09}.
\STATE \textbf{return}  $(\hat{\tau}_1,\hat{\theta}_1,\hat{\phi}_1) \ldots (\hat{\tau}_M,\hat{\theta}_M,\hat{\phi}_M)$.
\end{algorithmic}
\end{algorithm}

As per the bistatic \ac{RCS} estimates, they can be obtained given the per-path estimates $\{(\hat\theta_m,\hat\phi_m,\hat\tau_m,\hat\alpha_m)\}$ and the complex gains model in \eqref{eq:alpha_factor_local_paper} via
\begin{equation}
\label{eq:rcs_inversion_resp_paper}
\widehat{\sigma}^{\tt bi}_m
\;=\;
\frac{(4\pi)^3}{\lambda^2}\,
\frac{|\hat\alpha_m|^2\,\hat R_{{\tt t},m}^2\,\hat R_{{\tt r},m}^2}
{G_{\tt t}(\hat\phi_m)\,G_{\tt r}(\hat\theta_m)}.
\end{equation}
\noindent where $\hat R_{{\tt t},m}$ and $\hat R_{{\tt r},m}$ can be inferred from the estimated \ac{AoD}/\ac{AoA}/\ac{ToA} using standard hybrid localization methods \cite{geometry_pred}, and can be substituted in \eqref{eq:rcs_inversion_resp_paper} to infer the \ac{RCS} values.

\subsection{Extension to Doppler estimation}
\label{sec:methods-extension-doppler}
As discussed in Section~\ref{sec:system_model}, high-mobility scenarios induce Doppler shifts that manifest as time-dependent phase changes in the received signals. Therefore, accurately estimating these shifts is essential for determining the velocities of potential targets.
Let \(\pmb{G}_k(\pmb{f_D})\) be defined as in Eq.~\eqref{eq:CIR-freq-doppler}, incorporating both Doppler effects and path gains. Indeed, the methods must process the temporal evolution of the channel over multiple \ac{OFDM} symbols to incorporate Doppler estimation. Specifically, this requires estimating the channel over a series of sub-frames to capture the Doppler-induced phase shifts across time.
For the PARAMING algorithm, this adaptation allows for Doppler estimation with minimal structural changes. However, it necessitates operating over a sequence of sub-frames and ensuring that these sub-frames lie within a \ac{CPI}, where Doppler shifts influence the channel's temporal variation, but the channel remains stable enough for coherent processing.
Let \(\bar{\pmb{H}}_{n, p}\) denote the \ac{LS} channel estimate for the \(n^{\text{th}}\) subcarrier in the \(p^{\text{th}}\) sub-frame, as defined in \eqref{eq:estim_sub_frames}. By constructing the matrices from \eqref{method2_step1-1} through \eqref{method2_step1-3} and applying the PARAMING algorithm, the phase offsets for each target in the \(p^{\text{th}}\) sub-frame, denoted as \(\lbrace \hat{\delta}_m^{p} \rbrace_{m=1}^{M}\), can be estimated according to \eqref{eq:method2_step09}. Notably, the Hankel-block-Hankel-block-Hankel structure used for \ac{AoA}, \ac{AoD}, and \ac{ToA} estimation remains applicable. The temporal evolution of the phase offsets within \(\pmb{G}_k(\pmb{f_D})\) now embeds the Doppler shift information.
Assuming the estimation is conducted over \(\Tilde{K}_{\tt{p}}\) sub-frames, let \(\lbrace \hat{\delta}_m^{p} \rbrace_{p=1}^{\Tilde{K}_{\tt{p}}}\) denote the estimated phase offsets for the \(m^{\text{th}}\) target. These phase offsets can be expressed as
\begin{equation}
\hat{\delta}_m^{p} = \delta_m^0 + 2 \pi p K_{\tt{P}} T_o f_{D,m} + \Tilde{\varepsilon}_{m,p},
\end{equation}
\noindent where \(\delta_m^0\) represents the initial phase offset, \(f_{D,m}\) is the Doppler shift of the \(m^{\text{th}}\) target, and \(\Tilde{\varepsilon}_{m,p}\) denotes the phase estimation error. Since the phase evolves linearly with time in the presence of Doppler shifts, a simple $2$D regression on the estimated phase offsets suffices to extract the Doppler frequencies of all targets.

Similarly, extending the IFFT-\ac{C2VNN} algorithm to estimate Doppler requires modeling the temporal evolution of the channel over a \ac{CPI}. We stack the estimated \acp{CSI} from \(\Tilde{K}_{\tt p}\) successive sub-frames to form a $3$D input tensor (frequency \(\times\) space \(\times\) time). The \(2\)D complex-valued convolutions are then replaced by \(3\)D complex-valued convolutions so the network can learn spatio-temporal features that capture Doppler-induced phase progression, while the output head is augmented to predict four parameters per target: \ac{AoA}, \ac{AoD}, \ac{ToA}, and Doppler frequency. The temporal kernels encode inter-frame phase evolution, whereas the spatial-frequency kernels continue to extract angle and delay structure. To ensure robust performance, the training process involves generating datasets that span various Doppler shift scenarios, along with diverse \ac{SNR} conditions. This training strategy enables the model to generalize effectively across a wide range of operational settings.

As a lightweight alternative that avoids training a \(3\)D network, each sub-frame is processed independently with the original \(2\)D IFFT-\ac{C2VNN} to estimate \((\hat\theta_m^{p},\hat\phi_m^{p},\hat\tau_m^{p})\). Given these geometric parameters, the per-sub-frame complex amplitudes \(\hat\alpha_m^{p}\) are obtained by an \ac{LS} fit to the steering-delay dictionary across subcarriers. Doppler is then recovered from the slope of the unwrapped phases \(\angle \hat\alpha_m^{p}\) over time using the same \(1\)D linear regression employed in PARAMING. This alternative reuses the trained \(2\)D model and adds only light per-sub-frame \ac{LS} solves, with no need to retrain a \(3\)D convolutional network.
\section{Computational Complexity}
\label{sec:comput_complexity}
In this section, we comprehensively analyze the computational complexity of both IFFT-\ac{C2VNN} and PARAMING, quantifying it by the total number of additions and multiplications required for estimating the sensing parameters.

We begin by outlining the primary computational sub-blocks that significantly contribute to the complexity of PARAMING:
\begin{itemize}
    \item \ac{SVD} of $\widehat{\pmb{\mathcal{H}}}^{(1)}$ using the Golub-Reinsch algorithm \cite{ward1975combination}.
    \item Computation of $\pmb{T}$, as defined in \eqref{eq:method2_step4}, involving matrix multiplications among $\pmb{\bar{V}} \in \mathbb{C}^{K_{\tt{t}}K_{\tt{r}}(K_{N_{\tt{P}}}-1) \times M}$, $\pmb{\bar{U}}^H \in \mathbb{C}^{M \times M_{\tt{r}} M_{\tt{t}} M_{\tt{P}}}$, and the diagonal matrix $\pmb{\bar{\Sigma}}^{-1} \in \mathbb{C}^{M \times M}$.
    \item Eigenvalue decomposition of $\pmb{T} \in \mathbb{C}^{M \times M}$, using a QZ decomposition as $\pmb{T}$ has no specific structure.
    \item Calculation of \acp{ToA} using \eqref{eq:method2_step6} via a \ac{CORDIC} algorithm, which provides phase estimates of the eigenvalues.
    \item Estimation of $\widehat{\pmb{Y}}$ as per \eqref{eq:method2_step7}, followed by a 2D regression to estimate \acp{AoA} and \acp{AoD} as described in \eqref{eq:method2_step09}.
\end{itemize}
\noindent The computational complexity details of these sub-blocks are omitted due to lack of space. However, summing them yields the following total computational cost for PARAMING
{\normalsize
\begin{align}
    T_{\tt add} &= 9K_{\tt trp}^3 
    + 8M_{\tt tr}K_{\tt trp}^2 
    + 4M_{\tt tr}^2K_{\tt trp} 
    + M^3 \nonumber \\
    &\quad + M^2\left(M_{\tt tr} + 4N_{\tt cord} + 2N_{\tt P} - 4\right) \nonumber \\
    &\quad + M\big[M_{\tt tr}(K_{\tt trp} - 1) - 2N_{\tt cord} 
    + N_{\tt t}N_{\tt r}(N_{\tt P} + 2) \nonumber \\
    &\qquad - N_{\tt P} + 4\big] \nonumber ,
        \\[0.5em]
    T_{\tt mul} &= 9K_{\tt trp}^3 
    + 8M_{\tt tr}K_{\tt trp}^2 
    + 4M_{\tt tr}^2K_{\tt trp} 
    + M^3 \nonumber \\
    &\quad + M^2\left(M_{\tt tr} + 2N_{\tt P} + 3\right) \nonumber \\
    &\quad + M\left[M_{\tt tr}K_{\tt trp} + N_{\tt t}N_{\tt r}(N_{\tt P} + 3) + 9\right].
    \end{align}
}    
\noindent where \(T_{\tt{add}}\) and \(T_{\tt{mul}}\) denote the total number of additions and multiplications, respectively. Here, \(M_{\tt{tr}} = M_{\tt{t}}M_{\tt{r}}M_{\tt{P}}\), \(K_{\tt{trp}} = K_{\tt{t}}K_{\tt{r}}(K_{N_{\tt{P}}}-1)\), and \(N_{\tt{cord}}\) is the number of iterations for the \ac{CORDIC} algorithm.

For completeness, we also characterize the operation counts of two widely used grid-based baselines, namely \ac{DML} and \ac{MUSIC}. Let $D\triangleq N_{\tt r}N_{\tt t}N_{\tt P}$ be the ambient dimension of the vectorized per-subcarrier \ac{LS} channel, and let $(G_\theta,G_\phi,G_\tau)$ be the grid sizes for \ac{AoA}, \ac{AoD}, and \ac{ToA}. Denote the total number of voxels by $G\triangleq G_\theta G_\phi G_\tau$, and form the sample covariance from $K_{\tt snap}$ snapshots as $\pmb{R}=\frac{1}{K_{\tt snap}}\sum_{j=1}^{K_{\tt snap}}\pmb{\bar{H}}_j \, \pmb{\bar{H}}_j^{\!H}\in\mathbb{C}^{D\times D}$, where \(\pmb{\bar{H}}_j\) is the vectorized \ac{CSI} matrix for snapshot \(j\) (cf.\,\eqref{eq:method2_step0_1}). The covariance accumulation costs $T^{(\mathrm{cov})}_{\tt mul}=K_{\tt snap}D^2$ and $T^{(\mathrm{cov})}_{\tt add}=K_{\tt snap}D^2$.

The \ac{DML} score on a grid is $P_{\mathrm{DML}}(\theta,\phi,\tau)=\frac{\pmb{a}^{\!H}\pmb{R}\,\pmb{a}}{\pmb{a}^{\!H}\pmb{a}}$, where $\pmb{a}(\theta,\phi,\tau)=\pmb{c}(\tau)\otimes\pmb{a}_{\tt t}(\phi)\otimes\pmb{a}_{\tt r}(\theta)\in\mathbb{C}^{D}$. Each voxel evaluation requires one matrix-vector product $\pmb{R}\pmb{a}$ and two inner products. Summing over $G$ voxels and adding the covariance cost yields
\begin{equation}
\label{eq:comp_DML}
\begin{aligned}
T_{\tt mul}^{(\mathrm{DML})} &= K_{\tt snap}D^2 \;+\; G\,(D^2+2D),\\
T_{\tt add}^{(\mathrm{DML})} &= K_{\tt snap}D^2 \;+\; G\,(D^2 + D - 2).
\end{aligned}
\end{equation}

For \ac{MUSIC}, let $\pmb{R}=\pmb{U}\pmb{\Lambda}\pmb{U}^{\!H}$ and let $\pmb{U}_n\in\mathbb{C}^{D\times(D-s)}$ collect the noise eigenvectors for model order $s$. The spectrum is $P_{\text{MUSIC}}(\theta,\phi,\tau)=\frac{\pmb{a}^{\!H}\pmb{a}}{\pmb{a}^{\!H}\pmb{P}_n\pmb{a}}$, with $\pmb{P}_n=\pmb{U}_n\pmb{U}_n^{\!H}$. The costs consist of a Hermitian eigendecomposition and a per-voxel projection and norm. Using constants $(\gamma_{\tt evd},\eta_{\tt evd})$ for the leading $D^3$ terms of the eigensolver, the counts are
\begin{equation}
\label{eq:comp_MUSIC}
\begin{aligned}
T_{\tt mul}^{(\mathrm{MUSIC})} &= K_{\tt snap}D^2 \;+\; \gamma_{\tt evd}D^3 \;+\; G\,(D{-}s)(D{+}1),\\
T_{\tt add}^{(\mathrm{MUSIC})} &= K_{\tt snap}D^2 \;+\; \eta_{\tt evd}D^3 \;+\; G\,(D(D{-}s)-1).
\end{aligned}
\end{equation}
Equations~\eqref{eq:comp_DML}-\eqref{eq:comp_MUSIC} make explicit the dependence on the grid cardinality $G$, the ambient dimension $D$, the snapshot count $K_{\tt snap}$, and, for \ac{MUSIC}, the noise-subspace size $(D{-}s)$ and the eigendecomposition term.

\begin{figure}[!t]
\centering
\includegraphics[width=3.35in, trim=1.4mm 1.5mm 0mm 0mm, clip]{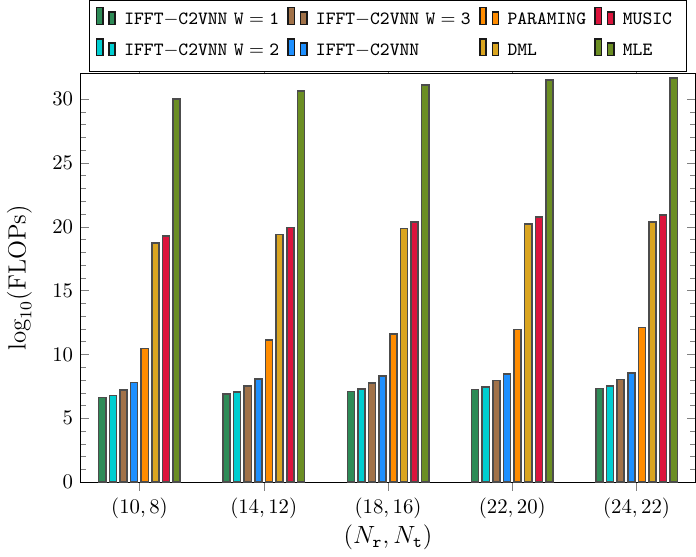}
\caption{%
Computational complexity in terms of operation counts. The plot includes IFFT-\ac{C2VNN} with window variants ($W$), PARAMING, and grid-based baselines (\ac{DML}, \ac{MUSIC}, and \ac{MLE}).}
\label{fig_complexity_Analysis}
\end{figure}

For IFFT-\ac{C2VNN}, the computational complexity is measured by counting the total operations within the architecture using the \ac{FLOPS} profiler in TensorFlow \cite{tensorflow2}. As outlined in Section~\ref{sec:ML_method}, we explore multiple window configurations around target peaks to reduce the computational cost.
To provide a benchmark, we also present the computational complexity of the grid-based \ac{MLE} method, as defined in \eqref{eq:complexity_MLE}. For comparability, we assume \( G_\alpha = 1 \), simulating a scenario where the path gain \(\alpha\) is known to the \ac{MLE}. We set \( G_\phi = G_\theta = 180 \), yielding a grid resolution of approximately \(1^\circ\) for \ac{AoA} and \ac{AoD} within a search range of \(-90^\circ\) to \(90^\circ\). For \ac{ToA}, we assign \( G_\tau = 128 \), which corresponds to a precision gain of $2$ beyond initial coarse estimates. For \ac{DML} and \ac{MUSIC}, we use a $0.01^\circ$ angular resolution in both $\theta$ and $\phi$ (i.e., $G_\theta=G_\phi=180/0.01=18{,}000$ points over $[-90^\circ,90^\circ]$) and a \ac{ToA} grid with $G_\tau = N_{\tt P}\times 5$, i.e., a precision factor of $5$.
In Fig.~\ref{fig_complexity_Analysis}, we present the computational complexity as a function of the number of transmit and receive antennas \((N_{\tt{r}}, N_{\tt{t}})\) in a scenario with three targets. The results clearly show that the \ac{MLE} requires significantly more operations, with approximately \(10^{22}\) times the operations of IFFT-\ac{C2VNN} and \(10^{19}\) times more than PARAMING.
Moreover, PARAMING generally incurs a higher complexity than the various configurations of IFFT-\ac{C2VNN}. Notably, as the system scale increases, the complexity of PARAMING escalates more rapidly than that of IFFT-\ac{C2VNN}, highlighting the latter’s suitability for resource-constrained scenarios.
The added \ac{DML} and \ac{MUSIC} curves quantify the cost of grid evaluations and, for \ac{MUSIC}, the eigendecomposition overhead. The dependence on $G$ and $D$ aligns with \eqref{eq:comp_DML}-\eqref{eq:comp_MUSIC} and makes explicit that finer angular and delay grids tighten mismatch at the expense of substantially higher runtimes and memory footprints.

In addition to the inference complexity, we evaluate the training complexity of IFFT-\ac{C2VNN}, focusing on the computational cost of forward and backward passes over multiple epochs. In complex-valued networks, each convolutional operation entails additional arithmetic due to the simultaneous processing of real and imaginary components.
For a network input \(\pmb{{\mathcal{H}}}_{\text{in}}^{1} \in \mathbb{C}^{(2W+1) \times N_{\tt{r}} N_{\tt{t}}}\), where \(W\) denotes the window size, each convolutional layer \(l = 1, \ldots, L\) comprises \(C_l\) complex filters \(\pmb{f}_{l}^{c_{l}} \in \mathbb{C}^{h_{f}^{l} \times w_{f}^{l}}\) with stride \(S_{l}\) and padding \(P_{l}\). The input and output dimensions at each layer are \(\pmb{{\mathcal{H}}}_{\text{in}}^{l} \in \mathbb{C}^{h^{l} \times w^{l} \times C_{l-1}}\) and \(\pmb{{\mathcal{H}}}_{\text{out}}^{l} \in \mathbb{C}^{h_{\text{out}}^{l} \times w_{\text{out}}^{l} \times C_{l}}\), where
\begin{equation}
    h_{\text{out}}^{l} = \frac{h^{l} + 2 P_{l} - h_{f}^{l}}{S_{l}} + 1, \quad w_{\text{out}}^{l} = \frac{w^{l} + 2 P_{l} - w_{f}^{l}}{S_{l}} + 1.
\end{equation}
The forward pass complexity per layer is
\begin{equation}
\label{eq:forward_cnn_complexity}
    \mathcal{C}_{\text{forward}} = \mathcal{O}\left( \sum_{l=1}^{L} 4 \, h_{\text{out}}^{l} w_{\text{out}}^{l} C_{l-1} h_{f}^{l} w_{f}^{l} C_{l} \right),
\end{equation}
where the factor of \(4\) accounts for real and imaginary components in complex-valued convolutions.

The backward pass, approximately twice as computationally intensive as the forward pass, includes error propagation, gradient calculation, and weight updates. With \(\mathcal{E}\) epochs and dataset size \(\mathcal{B}\), the total training complexity is
\begin{equation}
\label{eq:cnn_complexity_all}
    \mathcal{C}_{\text{training}} = \mathcal{O}\left( \sum_{l=1}^{L} 12 \, \mathcal{E} \mathcal{B} h_{\text{out}}^{l} w_{\text{out}}^{l} C_{l-1} h_{f}^{l} w_{f}^{l} C_{l} \right).
\end{equation}
It is important to note that this computational cost applies only during the training phase. For inference, only the forward pass is required, as defined in \eqref{eq:forward_cnn_complexity}, making real-time applications feasible with suitable hardware and software optimizations.

For the Doppler-enabled variant, \(\Tilde{K}_{\tt p}\) sub-frames are stacked to introduce a temporal dimension \(t\). Let
\(\pmb{\mathcal H}_{\text{in}}^{\,l}\in\mathbb{C}^{h^{l}\times w^{l}\times t^{l}\times C_{l-1}}\)
denote the input to layer \(l\), and use \(3\)D kernels of size \(h_f^{l}\times w_f^{l}\times t_f^{l}\) with temporal stride \(S_{t}^{l}\) and padding \(P_{t}^{l}\).
The temporal output length is
\[
t_{\text{out}}^{l} \;=\; \frac{t^{l} + 2P_{t}^{l} - t_f^{l}}{S_{t}^{l}} + 1.
\]
The \(3\)D forward-pass complexity becomes
\begin{equation}
\label{eq:forward_cnn_complexity_3d}
\mathcal{C}_{\text{forward}}^{(3\text{D})}
\;=\;
\mathcal{O}\!\left(
\sum_{l=1}^{L}
4\, h_{\text{out}}^{l} w_{\text{out}}^{l} t_{\text{out}}^{l}\, C_{l-1}\, h_f^{l} w_f^{l} t_f^{l}\, C_{l}
\right),
\end{equation}
and the corresponding training complexity is
\begin{equation}
\label{eq:cnn_complexity_all_3d}
\mathcal{C}_{\text{training}}^{(3\text{D})}
\;=\;
\mathcal{O}\!\left(
\sum_{l=1}^{L}
12\, \mathcal{E}\mathcal{B}\, h_{\text{out}}^{l} w_{\text{out}}^{l} t_{\text{out}}^{l}\, C_{l-1}\, h_f^{l} w_f^{l} t_f^{l}\, C_{l}
\right).
\end{equation}
Relative to \eqref{eq:forward_cnn_complexity}, the increase is linear in the temporal output length \(t_{\text{out}}^{l}\) and the kernel depth \(t_f^{l}\) for fixed spatial sizes. For typical \(\Tilde{K}_{\tt p}\) in the tens and compact temporal kernels, the additional cost remains modest with respect to the \(2\)D model and is compatible with real-time inference on modern accelerators.
\section{Numerical Evaluation}
\label{sec:simulations_new}
In this section, we present a comprehensive performance evaluation of the proposed PARAMING and IFFT-\ac{C2VNN} algorithms, alongside a comparison with state-of-the-art estimation methods, including Bartlett, MUSIC, Root-MUSIC, and \ac{DML} \cite{rootmusic, bartlett}. These methods represent well-established baselines for parameter estimation and were specifically adapted to our communication-centric \ac{ISAC} context. By incorporating these methods into our analysis, we aim to highlight the comparative advantages of the proposed approaches under varied noise conditions.
The simulation setup adheres to realistic \ac{ISAC} system configurations, as summarized in Table~\ref{table:simulation_parameters}. 
Unless otherwise stated, all Monte-Carlo experiments use \(M{=}3\) targets.
\begin{table}[!t]
\centering
\caption{Simulation Parameters}
\label{table:simulation_parameters}
\begin{tabular}{@{}ll@{}}
\toprule
\textbf{Parameter}    & \textbf{Value}     \\ 
\midrule
Number of transmit antennas $(N_{\tt{t}})$ & $8$ \cite{10437321,10158439} \\
Number of receive antennas $(N_{\tt{r}})$ & $10$ \cite{10437321} \\
Carrier frequency $(f_c)$        &  $28 \ \si{\GHz}$ \cite{10158439}             \\
Antenna spacing $\left(d_{\tt{r}} = d_{\tt{t}} = \lambda / 2\right)$               & $\SI{0.53}{\centi\meter} $            \\
Number of subcarriers $(N_{\tt{P}})$               & $64$ \cite{9456022}            \\
Number of \ac{OFDM} symbols per sub-frame $(K_{\tt{P}})$             & $10$ \cite{9456022}           \\
Number of sub-frames $(\Tilde{K}_{\tt{p}})$             & $4$ \cite{germany5g}\\
Subcarrier spacing $(\Delta_f)$               & $\SI{960}{\kilo\hertz}$ \cite{10330696}           \\
\ac{OFDM} symbol duration $(T_o)$              & $\SI{1.3}{\micro\second}$            \\
Temporal resolution $\left(\Delta_{\tt{t}} = \frac{1}{N_{\tt{P}} \Delta_f}\right)$               & $\SI{16.27}{\nano\second}$            \\
\bottomrule
\end{tabular}
\end{table}
The \ac{CIR} model used in our simulations, as defined in Eq.~\eqref{eq:CIR}, accounts for the superposition of reflections from multiple scatterers, including both targets and environmental clutter. A refined classification of scatterers, achieved by estimating the sensing parameters across multiple coherence time intervals or employing advanced \ac{ML} techniques, can enhance the distinction between these entities by enabling the identification of scatterer dynamics and facilitating the discrimination between static clutter and moving targets.
Unless otherwise specified, we adopt Swerling-I type priors for per-path amplitudes in simulations. The proposed methods are compatible with distinct priors for targets and clutter (e.g., different distributions for the bistatic \ac{RCS} $\sigma^{\tt{bi}}_{m}$ or directly for the complex gain $\alpha_{m}$) without modifying the estimators.

For the benchmarked grid-based methods, namely Bartlett, MUSIC, and \ac{DML}, predefined search grids were employed to estimate the sensing parameters. To balance estimation accuracy and computational feasibility, the grid resolution was set to \(0.05^\circ\) for \ac{AoA} and \ac{AoD}, and \(\Delta_{\tt{t}}/5\) for \ac{ToA}. These resolutions ensure fair comparisons with the proposed methods while keeping computational complexity manageable. Additionally, the \ac{CRB}, which serves as the theoretical performance bound, is discussed in \textbf{Appendix~\ref{app:appendix-CRB}}. For the PARAMING method, the sub-array dimensions were set to \(M_{\tt{t}} =\lfloor \frac{N_{\tt{t}}}{2} \rceil\), \(M_{\tt{r}} =\lfloor \frac{N_{\tt{r}}}{2} \rceil\), and the sub-\ac{OFDM} symbol dimension to \(M_{\tt{P}} =\lfloor \frac{N_{\tt{P}}}{2} \rceil\).

\begin{figure}[!ht]
\centering
\includegraphics[width=3.35in, trim=10mm 0mm 0mm 0mm, clip]{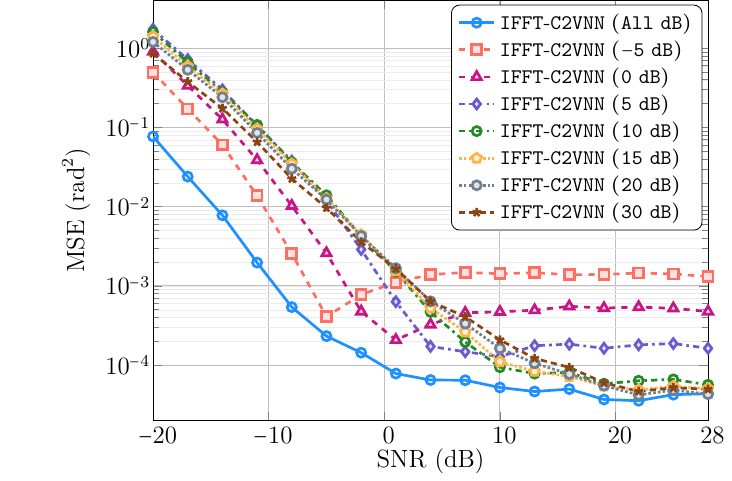}
\caption{\ac{AoA} \ac{MSE} (\si{\radian\squared}) vs \ac{SNR}. \ac{AoD} and \ac{ToA} follow the same trend and are omitted for brevity.}
\label{fig_aoa_snr_analysis}
\end{figure}

To identify a robust training strategy for IFFT-\ac{C2VNN} under varying noise conditions, we compare specialist models trained at fixed \ac{SNR} levels \(\{-5,0,5,10,15,20,30\}\,\si{\decibel}\) with a single mixed-\ac{SNR} model trained on data whose \ac{SNR} is drawn uniformly from a broad range \([-5,40]\,\si{\decibel}\). All models share the same architecture, loss, and optimizer (Adam with learning rate \(10^{-4}\)). We use the same batch size (128), number of epochs (300), and geometry distribution, so the only difference is the \ac{SNR} distribution used during training. Fig.~\ref{fig_aoa_snr_analysis} reports the \ac{AoA} \ac{MSE} versus \ac{SNR}. The \ac{AoD} and \ac{ToA} curves follow the same trend. The mixed-\ac{SNR} model closely follows the lower envelope of the specialist curves across the entire sweep. Specialist models perform best near their training \ac{SNR} and degrade when evaluated far from it. In particular, low-\ac{SNR} specialists plateau at high \ac{SNR}, and high-\ac{SNR} specialists lose robustness at very low \ac{SNR}. Because the deployment \ac{SNR} is unknown and time varying, we adopt mixed-\ac{SNR} training by default.
We also evaluated a variant of IFFT-\ac{C2VNN} with \(W\,=\,2\) to assess the effect of restricting the input \ac{CSI} to peak regions, which reduces noise and improves convergence. Monte Carlo experiments were conducted using independently generated channel realizations, random scatterer locations, and \ac{SNR} values spanning \SI{-20}{\decibel} to \SI{31}{\decibel}.
\begin{figure}[!ht]
    \centering
    \includegraphics[width=3.35in, trim=10mm 0mm 0mm 0mm, clip]{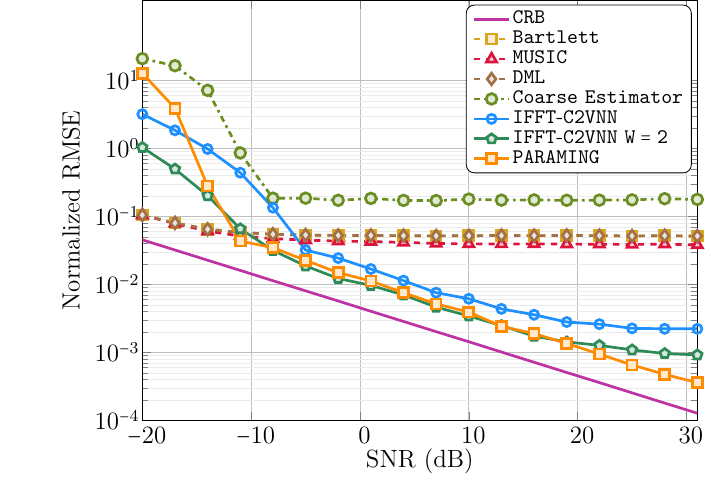}
\caption{Normalized \ac{RMSE} of \ac{ToA} estimation for the proposed methods and benchmarked approaches, compared against the \ac{CRB}.}
\label{fig:toa_performance_comparison}
\end{figure}

Fig.~\ref{fig:toa_performance_comparison} illustrates the \ac{ToA} estimation performance in terms of \ac{RMSE} normalized by the system's temporal resolution \(\Delta_{\tt{t}}\). Among the benchmarked methods, Bartlett, MUSIC, and \ac{DML} exhibit identical performance under the considered scenario.
The performance of the grid-based methods saturates at high \ac{SNR} values due to the predefined grid resolution \(\Delta_{\tt{t}}/5\), which imposes a quantization floor, and while refining the grid would reduce this floor, it would incur substantially higher computational and memory costs and would distort fairness relative to the proposed methods. 
By contrast, the proposed methods, PARAMING and IFFT-\ac{C2VNN}, demonstrate superior performance, estimating effectively across both low and high \ac{SNR} levels. PARAMING achieves the closest alignment to the \ac{CRB} at high \ac{SNR}, reflecting its enhanced precision. IFFT-\ac{C2VNN}, trained over a broad \ac{SNR} range, exhibits strong robustness, while its \(W=2\) variant further improves accuracy by focusing on peak regions of the \ac{CSI} input.
\begin{figure*}[!t]
    \centering
    \begin{subfigure}[t]{3.5in}
        \centering
        \includegraphics[width=3.35in, trim=11mm 0mm 0mm 0mm, clip]{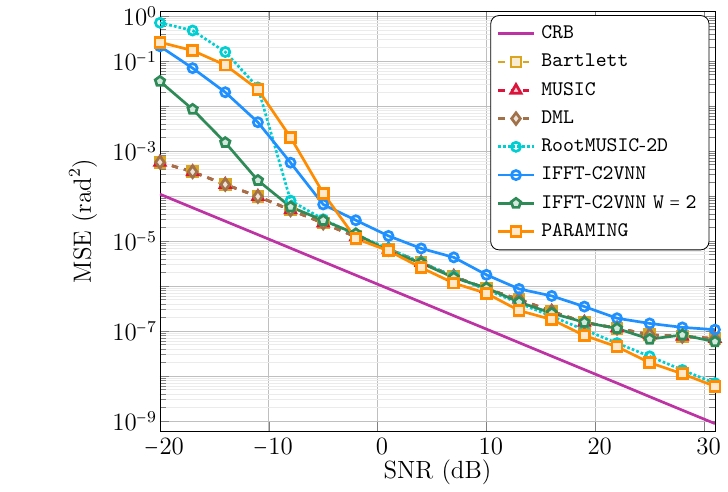}
        \caption{\ac{AoA} estimation performance.}
        \label{fig:aoa_comparison}
    \end{subfigure}
    \hfill
    \begin{subfigure}[t]{3.5in}
        \centering
        \includegraphics[width=3.35in, trim=11mm 0mm 0mm 0mm, clip]{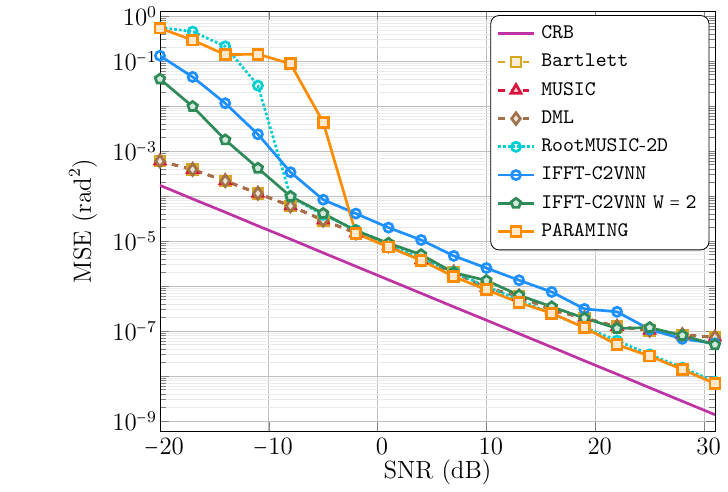}
        \caption{\ac{AoD} estimation performance.}
        \label{fig:aod_comparison}
    \end{subfigure}
    \caption{Comparison of \ac{AoA} and \ac{AoD} estimation performance in terms of \ac{MSE} across \ac{SNR} levels.}
    \label{fig:aoad_performance_comparison}
\end{figure*}

The performance comparison for \ac{AoA} and \ac{AoD} estimation is presented in Fig.~\ref{fig:aoad_performance_comparison}, highlighting the \ac{MSE} across the same range of \ac{SNR} levels. Similar to the \ac{ToA} case, Bartlett, MUSIC, and \ac{DML} exhibit nearly identical performance, constrained by the \(0.05^\circ\) grid resolution. RootMUSIC-2D, despite only estimating \ac{AoA} and \ac{AoD} and leveraging data from all subcarriers, achieves reasonable performance but does not surpass PARAMING. Nevertheless, it demonstrates an earlier waterfall region, providing an advantage in low-\ac{SNR} scenarios. Finally, PARAMING achieves the closest alignment to the \ac{CRB}, while IFFT-\ac{C2VNN} shows strong robustness but stagnates at very high \ac{SNR} levels. The \(W=2\) variant of IFFT-\ac{C2VNN} further improves accuracy, reinforcing the effectiveness of the proposed methods for sensing parameter estimation.
When compared to the \ac{CRB}, all compared methods, including both proposed methods, exhibit sub-optimal performance across all sensing parameters. For instance, in the case of \ac{ToA} estimation, achieving a normalized \ac{RMSE} of $10^{-2}$ requires an \ac{SNR} approximately $\SI{9}{\decibel}$ higher than the \ac{CRB}. A similar performance gap is observed for the \ac{AoA} and \ac{AoD} estimations, as shown in Fig.~\ref{fig:aoad_performance_comparison}.

To evaluate the computational efficiency, we measured the execution times of the proposed and benchmarked methods on a \ac{HPC} cluster equipped with AMD EPYC $7742$ $64$-Core Processors operating at $\SI{2.25}{\giga\hertz}$. Each node consists of $128$ CPU cores and $480$ GB of memory; however, the jobs were executed on nodes specifically configured with $32$ GB of memory to simulate realistic computational constraints. Table~\ref{table:latency_comparison_combined} summarizes the latency results, averaged over $300$ Monte Carlo trials.
\begin{table}[!t]
\centering
\caption{Execution time of the proposed and benchmarked methods.}
\label{table:latency_comparison_combined}
\begin{tabular}{@{}lcc@{}}
\toprule
\textbf{Category} & \textbf{Method} & \textbf{Execution Time (s)} \\ 
\midrule
\multirow{2}{*}{\textbf{Proposed Methods}} 
    & IFFT-\ac{C2VNN} & $\mathbf{0.06}$ \\ 
    & PARAMING & $0.26$ \\ 
\midrule  
\multirow{4}{*}{\textbf{Benchmarked Methods}} 
    & Bartlett & $96.85$ \\ 
    & MUSIC & $104.07$ \\ 
    & RootMUSIC-2D & $38.81$ \\ 
    & DML & $95.61$ \\ 
\bottomrule
\end{tabular}
\end{table}
Among the compared methods, IFFT-\ac{C2VNN} achieves the lowest latency of $\mathbf{0.06}$ seconds, followed by PARAMING at $\mathbf{0.26}$ seconds. These results underscore the computational efficiency of the proposed methods, with latencies remaining well within the acceptable range for real-time \ac{ISAC} applications, particularly when leveraging hardware accelerators such as \acp{GPU} or \acp{FPGA}. By contrast, the grid-based benchmarked methods exhibit significantly higher latencies due to their reliance on exhaustive grid searches over fine-grained grids, which impose a substantial computational burden. Furthermore, these methods necessitate large dictionaries of steering vectors, thereby introducing scalability challenges and considerable space complexity. 
RootMUSIC-2D, which estimates only \ac{AoA} and \ac{AoD}, achieves a substantially lower latency compared to other grid-based methods. However, its latency still far exceeds that of the proposed methods. Moreover, extending RootMUSIC to a $3$D variant for simultaneous \ac{AoA}, \ac{AoD}, and \ac{ToA} estimation proves computationally infeasible due to the exponential increase in complexity associated with higher-dimensional polynomial-based estimation.
In summary, the proposed methods not only achieve superior estimation accuracy but also attain markedly lower latency than the benchmarks, demonstrating their suitability for efficient and scalable communication-centric \ac{ISAC}.

\begin{figure}[!ht]
\centering
\includegraphics[width=3.35in, trim=11mm 0mm 0mm 0mm, clip]{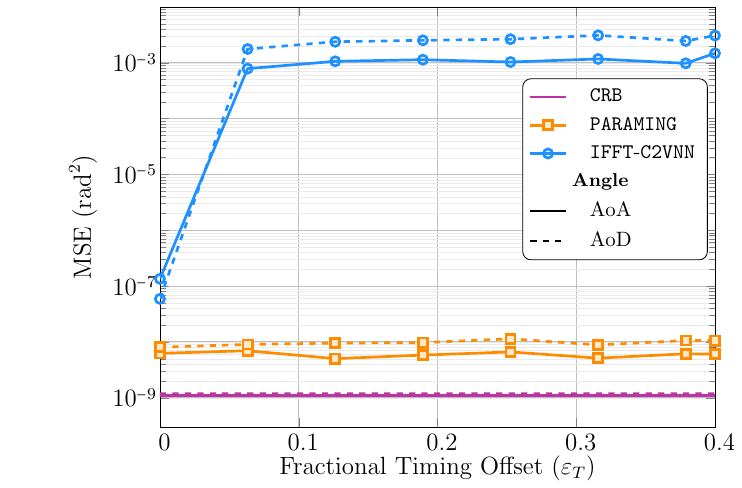}
\caption{\ac{AoA} and \ac{AoD} estimation error versus \ac{TO} \((\tau_{\mathrm{off}}=\varepsilon_T T_o)\).}
\label{fig:synchro_to_frac_to_aoa_aod}
\end{figure}
We further assess robustness to \ac{TO} by modeling a deterministic offset \(\tau_{\mathrm{off}}=\varepsilon_T T_o\) with \(\varepsilon_T\in[0,0.4]\). Fig.~\ref{fig:synchro_to_frac_to_aoa_aod} reports the angular accuracy under this condition. PARAMING remains essentially invariant across \(\varepsilon_T\) for \ac{AoA}/\ac{AoD}, since the eigendecomposition step in \eqref{eq:method2_step4} operates on \(\{e^{-j2\pi n\Delta_f(\tau_m+\tau_{\mathrm{off}})}\}_n\) and cancels the common \ac{TO}, thereby preserving the spatial phases. This invariance can introduce a systematic bias in \ac{ToA} if \(\tau_{\mathrm{off}}\) is not explicitly compensated. In contrast, IFFT-\ac{C2VNN} exhibits a gradual degradation as \(\varepsilon_T\) increases, because the \ac{TO} alters the per-subcarrier phases and induces a distribution shift relative to the training input. In practice, this sensitivity can be mitigated by augmenting the training set with synthetic \acp{TO} or by applying a lightweight pre-compensation before inference.

\begin{figure}[!ht]
    \centering
    \includegraphics[width=3.3in, trim=5mm 0mm 0mm 0mm, clip]{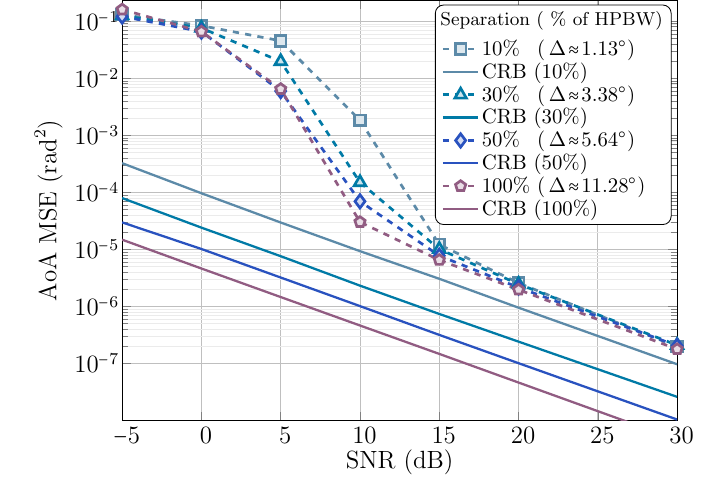}
    \caption{\ac{AoA} MSE versus \ac{SNR} for an experiment with two targets where the angular separations are expressed as a fraction of the receive array \ac{HPBW}. Solid lines: \ac{CRB} for each separation. Dashed lines with markers: PARAMING \ac{MSE}.}
    \label{fig:snr_sep}
\end{figure}

To assess separability in close proximity scenarios, we consider an experiment with two targets in which the \acp{AoA} are separated by a fixed fraction of the receive array \ac{HPBW}. For the \(N_{\tt r}{=}10\) \ac{ULA} employed, \(\mathrm{HPBW}\!\approx\!11.28^\circ\), so the fractions \(\{10,30,50,100\}\%\) correspond to \(\{1.13^\circ,3.38^\circ,5.64^\circ,11.28^\circ\}\). Fig.~\ref{fig:snr_sep} reports the \ac{AoA} \ac{MSE} versus \ac{SNR} together with the corresponding \ac{CRB} for each separation and in which two trends emerge. First, as separation decreases, the \ac{CRB} shifts upward, reflecting the growing difficulty of discrimination at fixed \ac{SNR}. For example, attaining the same \ac{MSE} at \(50\%\) of \ac{HPBW} requires several additional \si{\decibel} relative to the \(100\%\) case (about \(4\) \si{\decibel} in our setting). Second, PARAMING achieves super-resolution below the \ac{HPBW} and exhibits a threshold "waterfall" behavior. Once the \ac{SNR} exceeds a separation-dependent breakpoint, the estimation error decreases and the performance curves converge to a nearly separation-agnostic trend. At high \ac{SNR}, the remaining gap to the \ac{CRB} dominates and becomes essentially independent of the separation. Similar qualitative trends hold for \ac{AoD} and \ac{ToA} and are omitted for brevity.

Although presented independently, PARAMING and IFFT-\ac{C2VNN} are complementary. A practical deployment can run IFFT-\ac{C2VNN} by default and invoke PARAMING only when a simple gate (for example \ac{NN} uncertainty or a large model-fit residual) flags doubt. Another option is to run PARAMING once per \ac{CPI} to resolve subpaths and then use a multi-head IFFT-\ac{C2VNN} to track $(\theta,\phi,\tau)$ across frames. The design and validation of such system-level synergies are left to future work.

\begin{figure}[!t]
\centering
\includegraphics[width=3.35in, trim=11mm 0mm 0mm 0mm, clip]{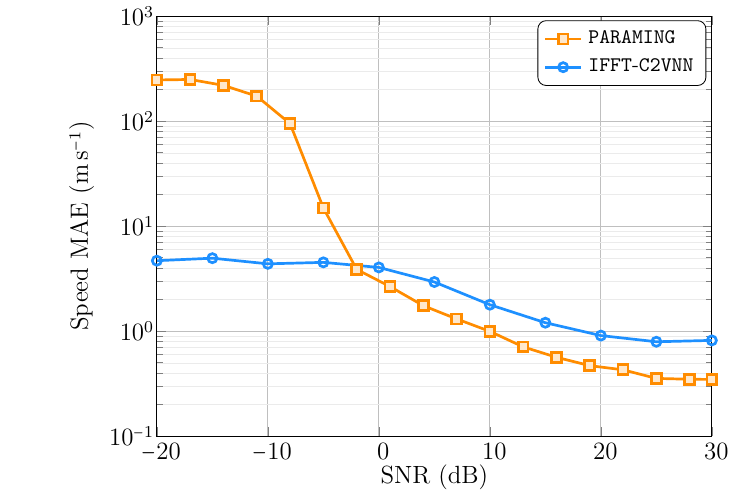}
\caption{Performance of the proposed methods for speed estimation, measured in \ac{MAE}.}
\label{fig:doppler_estim_performance}
\end{figure}

In addition to the joint estimation of \ac{AoA}, \ac{AoD}, and \ac{ToA}, we extended the proposed PARAMING and IFFT-\ac{C2VNN} methods to include Doppler frequency estimation, as detailed in Section~\ref{sec:methods-extension-doppler}. Simulations were conducted to evaluate their performance, considering target velocities uniformly distributed up to $\SI{30}{\meter \per \second}$, corresponding to Doppler shifts of up to $\pm \SI{2.8}{\kilo\hertz}$.
The Doppler estimation performance, measured in terms of \ac{MAE}, is presented in Fig.~\ref{fig:doppler_estim_performance}. The results demonstrate that both methods reliably estimate Doppler frequencies, with PARAMING achieving superior accuracy at high \ac{SNR} levels. Specifically, PARAMING attains an \ac{MAE} as low as $\SI{0.3}{\meter \per \second}$, compared to $\SI{0.8}{\meter \per \second}$ for IFFT-\ac{C2VNN}. These findings reaffirm the potential of both methods for high-resolution sensing parameter estimation.
\section{Conclusion}
\label{sec:conclusion}
In this paper, we introduced two approaches, IFFT-\ac{C2VNN} and PARAMING, for joint sensing-parameter estimation in a communication-centric bistatic \ac{ISAC} configuration, where a passive radar leverages downlink communication signals from existing infrastructure to estimate the parameters of multiple targets in the environment. PARAMING is a parametric method that exploits system-specific characteristics, such as array geometries, to extract the sensing parameters from a Hankel matrix constructed from the estimated \ac{CSI}. In contrast, IFFT-\ac{C2VNN} uses a \ac{DL}-based convolutional architecture to process the estimated channel for parameter estimation.
Performance results underscore the effectiveness of the proposed methods in enhancing the sensing capabilities within communication-centric \ac{ISAC} systems while also reducing the computational complexity. Future research directions include extending the estimation framework to fully capture the multidimensional nature of the problem by incorporating both azimuth and elevation angles, \acp{ToA}, Doppler shifts, and fully polarimetric per-path complex gains. This extension should also consider realistic factors such as local scattering, path loss, synchronization offsets, and cross-polarization effects to ensure robustness and adaptability in diverse scenarios.
While this work focuses on a single passive radar, a natural next step is multi-radar cooperative sensing. Key challenges include scalability, time/frequency synchronization, inter-radar association, and information-fusion strategies.

\appendices
\label{sec:appendix}
\section{Cram\'er-Rao Bound Expressions}
\label{app:appendix-CRB}
\small
The \ac{FIM} is given as follows
\begin{equation}
    \label{eq:FIM}
    \begin{split}
    \pmb{\Gamma}
   & \triangleq
    \mathbb{E} \Big[\frac{\partial \mathcal{L}(\pmb{\xi} )}{\partial \pmb{\xi} }  \frac{\partial\mathcal{L}(\pmb{\xi} )}{\partial \pmb{\xi} }^T \Big],
    \end{split}
\end{equation}
where $\pmb{\xi}$ is the vector of unknown parameters, namely $\pmb{\xi} = \begin{bmatrix}
    \sigma & \pmb{\Theta} & \pmb{\Phi} & \pmb{\tau} & \bar{\pmb{\alpha}} & \tilde{\pmb{\alpha}}
\end{bmatrix}$,
where $\bar{\pmb{\alpha}}$ represents the real-part of $\pmb{\alpha}$ 
and $\tilde{\pmb{\alpha}}$ represents the imaginary-part of $\pmb{\alpha}$.
Furthermore, $\mathcal{L}(\pmb{\xi} )$ is the log-likelihood of the model, i.e. $\mathcal{L}(\pmb{\xi} ) = \log f(\pmb{\mathcal{Y}})$ and $ f(\pmb{\mathcal{Y}})$ is the \ac{PDF} of the observed data defined in \eqref{eq:likelihood}.
The \ac{FIM} is partitioned according to the unknown variables, i.e. for any two parameter quantities, $\pmb{\Gamma}_{\pmb{a},\pmb{b}} =  \mathbb{E} [\frac{\partial \mathcal{L}(\pmb{\xi} )}{\partial \pmb{a} }  \frac{\partial\mathcal{L}^T(\pmb{\xi} )}{\partial \pmb{b} } ]$. Note that it is easy to see $\pmb{\Gamma}_{\sigma,\sigma} = \frac{N_{\tt{r}} N_{\tt{p}}K_{\tt{p}}}{\sigma^4} $ and $\pmb{\Gamma}_{\sigma,\pmb{\Theta}}
=
\pmb{\Gamma}_{\sigma,\pmb{\Phi}}
=
\pmb{\Gamma}_{\sigma,\pmb{\tau}}
=
\pmb{\Gamma}_{\sigma,\bar{\pmb{\alpha}}}
=
\pmb{\Gamma}_{\sigma,\tilde{\pmb{\alpha}}}
=
\pmb{0}^T$
Now, denoting $\pmb{\Xi}_i = \pmb{a}_r(\theta_i) \pmb{a}_t^T(\phi_i)$, $\pmb{\Xi}_i^r = \pmb{d}_r(\theta_i) \pmb{a}_t^T(\phi_i)$ and $\pmb{\Xi}_i^t = \pmb{a}_r(\theta_i) \pmb{d}_t^T(\phi_i) $, where $\pmb{d}_r(\theta) = \frac{\partial \pmb{a}_r(\theta)}{\partial \theta }$ and $\pmb{d}_t(\phi) = \frac{\partial \pmb{a}_t(\phi)}{\partial \phi }$ are the partial derivatives of the receive and transmit steering vectors with respect to $\theta$ and $\phi$, respectively.
To this end, we summarize the \ac{FIM} block-matrices appearing in \eqref{eq:FIM} as follows. 
First, we compute all second-order partial derivatives whenever $\pmb{\Theta}$ appears, i.e.
\begin{align*}
 [\pmb{\Gamma}_{\pmb{\Theta},\pmb{\Theta}}]_{i,j}
    &=
    \frac{2}{\sigma^2}
    \sum\limits_{n,k}
    \Re
    \Big(
     \pmb{s}_{n,k}^H
    [\alpha_i c_n(\tau_i) \pmb{\Xi}_i^r]^H
    [\alpha_j c_n(\tau_j) \pmb{\Xi}_j^r]
    \pmb{s}_{n,k}
    \Big), \\ 
    [\pmb{\Gamma}_{\pmb{\Theta},\pmb{\Phi}}]_{i,j}
    &=
    \frac{2}{\sigma^2}
    \sum\limits_{n,k}
    \Re
    \Big(
    \pmb{s}_{n,k}^H
    [\alpha_i c_n(\tau_i)  \pmb{\Xi}_i^r]^H
    [\alpha_j c_n(\tau_j)  \pmb{\Xi}_j^t]
    \pmb{s}_{n,k}
    \Big), \\
    [\pmb{\Gamma}_{\pmb{\Theta},\pmb{\tau}}]_{i,j}
    &=
    \frac{2}{\sigma^2}
    \sum\limits_{n,k}
    \Re
    \Big(
    \pmb{s}_{n,k}^H
    [\alpha_i c_n(\tau_i)  \pmb{\Xi}_i^r]^H
    [\alpha_j d_n(\tau_j)  \pmb{\Xi}_j]
    \pmb{s}_{n,k}
    \Big), \\
    [\pmb{\Gamma}_{\pmb{\Theta},\bar{\pmb{\alpha}}}]_{i,j}
    &=
    \frac{2}{\sigma^2}
    \sum\limits_{n,k}
    \Re
    \Big(
    \pmb{s}_{n,k}^H
    [\alpha_i c_n(\tau_i)\pmb{\Xi}_i^r]^H
    [c_n(\tau_j) \pmb{\Xi}_j]
    \pmb{s}_{n,k}
    \Big), \\
    [\pmb{\Gamma}_{\pmb{\Theta},\tilde{\pmb{\alpha}}}]_{i,j}
    &=
    \frac{2}{\sigma^2}
    \sum\limits_{n,k}
    \Re
    \Big(
    \pmb{s}_{n,k}^H
    [\alpha_i c_n(\tau_i) \pmb{\Xi}_i^r]^H
    [j c_n(\tau_j) \pmb{\Xi}_j]
    \pmb{s}_{n,k}
    \Big),
\end{align*}
where $d_n(\tau) = \frac{\partial c_n(\tau)}{\partial \tau}$. Then, we compute all second-order partial derivatives whenever $\pmb{\Phi}$ appears, i.e.
\begin{align*}
    [\pmb{\Gamma}_{\pmb{\Phi},\pmb{\Phi}}]_{i,j}
    &=
    \frac{2}{\sigma^2}
    \sum\limits_{n,k}
    \Re
    \Big(
    \pmb{s}_{n,k}^H
    [\alpha_i c_n(\tau_i) \pmb{\Xi}_i^t]^H
     [\alpha_j c_n(\tau_j) \pmb{\Xi}_j^t]
    \pmb{s}_{n,k}
    \Big), \\ 
    [\pmb{\Gamma}_{\pmb{\Phi},\pmb{\tau}}]_{i,j}
    &=
    \frac{2}{\sigma^2}
    \sum\limits_{n,k}
    \Re
    \Big(
    \pmb{s}_{n,k}^H
    [\alpha_i c_n(\tau_i) \pmb{\Xi}_i^t]^H
     [\alpha_j d_n(\tau_j) \pmb{\Xi}_j]
    \pmb{s}_{n,k}
    \Big), \\ 
    [\pmb{\Gamma}_{\pmb{\Phi},\bar{\pmb{\alpha}}}]_{i,j}
    &=
    \frac{2}{\sigma^2}
    \sum\limits_{n,k}
    \Re
    \Big(
    \pmb{s}_{n,k}^H
    [\alpha_i c_n(\tau_i) \pmb{\Xi}_i^t]^H
     [ c_n(\tau_j) \pmb{\Xi}_j]
    \pmb{s}_{n,k}
    \Big), \\ 
    [\pmb{\Gamma}_{\pmb{\Phi},\tilde{\pmb{\alpha}}}]_{i,j}
    &=
    \frac{2}{\sigma^2}
    \sum\limits_{n,k}
    \Re
    \Big(
    \pmb{s}_{n,k}^H
    [\alpha_i c_n(\tau_i) \pmb{\Xi}_i^t]^H
     [ j c_n(\tau_j) \pmb{\Xi}_j]
    \pmb{s}_{n,k}
    \Big),
\end{align*}
\!
Following the above expressions, we compute all \ac{FIM} partial derivatives where $\pmb{\tau}$ appears
\begin{align*}
    [\pmb{\Gamma}_{\pmb{\tau},\pmb{\tau}}]_{i,j}
    &=
    \frac{2}{\sigma^2}
    \sum\limits_{n,k}
    \Re
    \Big(
    \pmb{s}_{n,k}^H
    [\alpha_i d_n(\tau_i) \pmb{\Xi}_i]^H
     [\alpha_j d_n(\tau_j) \pmb{\Xi}_j]
    \pmb{s}_{n,k}
    \Big), \\ 
    [\pmb{\Gamma}_{\pmb{\tau},\bar{\pmb{\alpha}}}]_{i,j}
    &=
    \frac{2}{\sigma^2}
    \sum\limits_{n,k}
    \Re
    \Big(
    \pmb{s}_{n,k}^H
    [\alpha_i d_n(\tau_i) \pmb{\Xi}_i]^H
     [ c_n(\tau_j) \pmb{\Xi}_j]
    \pmb{s}_{n,k}
    \Big), \\ 
    [\pmb{\Gamma}_{\pmb{\tau},\tilde{\pmb{\alpha}}}]_{i,j}
    &=
    \frac{2}{\sigma^2}
    \sum\limits_{n,k}
    \Re
    \Big(
    \pmb{s}_{n,k}^H
    [\alpha_i d_n(\tau_i) \pmb{\Xi}_i]^H
     [ j c_n(\tau_j) \pmb{\Xi}_j]
    \pmb{s}_{n,k}
    \Big),
\end{align*}
Next, we compute all partial derivatives where $\bar{\pmb{\alpha}}$ appears
\begin{align*}
[\pmb{\Gamma}_{\bar{\pmb{\alpha}},\bar{\pmb{\alpha}}}]_{i,j}
    &
    =
    \frac{2}{\sigma^2}
    \sum\limits_{n,k}
    \Re
    \Big(
    \pmb{s}_{n,k}^H
    [c_n(\tau_i) \pmb{\Xi}_i]^H
     [c_n(\tau_j) \pmb{\Xi}_j]
    \pmb{s}_{n,k}
    \Big), \\
    [\pmb{\Gamma}_{\bar{\pmb{\alpha}},\tilde{\pmb{\alpha}}}]_{i,j}
    &
    =
    \frac{2}{\sigma^2}
    \sum\limits_{n,k}
    \Re
    \Big(
    \pmb{s}_{n,k}^H
    [c_n(\tau_i) \pmb{\Xi}_i]^H
     [j c_n(\tau_j) \pmb{\Xi}_j]
    \pmb{s}_{n,k}
    \Big),
\end{align*}
Then we compute all partial derivatives where $\tilde{\pmb{\alpha}}$ appears
\begin{equation*}
    [\pmb{\Gamma}_{\tilde{\pmb{\alpha}},\tilde{\pmb{\alpha}}}]_{i,j} 
    =
    \frac{2}{\sigma^2}
    \sum\limits_{n,k}
    \Re
    \Big(
    \pmb{s}_{n,k}^H
    [j c_n(\tau_i) \pmb{\Xi}_i]^H
     [j c_n(\tau_j) \pmb{\Xi}_j]
    \pmb{s}_{n,k}
    \Big).
\end{equation*}
Now, the \ac{CRB} for the parameters of interest (i.e, \ac{AoA}, \ac{AoD}, \ac{ToA}) is obtained as follows
\begin{equation}
    \begin{split}
     \CRB (\pmb{\Theta}) 
    &=
   [ \pmb{\Gamma}^{-1} ]_{2:(M+1),2:(M+1)}  \\
    \CRB (\pmb{\Phi}) 
    &=
    [\pmb{\Gamma}^{-1}]_{(M+2):(2M+1),(M+2):(2M+1)} \\
    \CRB (\pmb{\tau})
    &=
    [\pmb{\Gamma}^{-1}]_{(2M+2):(3M+1),(2M+2):(3M+1)}
    \end{split}
\end{equation}
\normalsize
\bibliographystyle{ieeetr}
\bibliography{references_ha}{}

\end{document}